\documentclass[prl,floatfix]{revtex4} 

\usepackage{graphicx}
\usepackage{psfrag}

\begin{document}

\newcommand{\dfrac}[2]{{\displaystyle\frac{#1}{#2}}}
\newcommand{\tfrac}[2]{{\textstyle\frac{#1}{#2}}}

\def\t{temperature }
\def\tn{temperature}
\def\dm{density matrix }
\def\dmn{density matrix}
\def\ady{adiabaticity }
\def\adyn{adiabaticity}
\def\ad{adiabatic }
\def\adn{adiabatic}
\def\de{decoherence }
\def\den{decoherence}
\def\sy{system }
\def\syn{system}
\def\wv{wavefunction }
\def\wvn{wavefunction}
\def\wvs{wavefunctions }
\def\wvsn{wavefunctions}
\def\hn{Hamiltonian}
\def\h{Hamiltonian }
\def\pl{$(\phi ^{ext}_1,\phi ^{ext}_2)$}
\def\px{$\phi^{ext}$}
\def\sy{system }
\def\syn{system}
\def\sq{SQUID }
\def\sqn{SQUID}

\title{%
\hbox to\hsize{\normalsize\rm 
\hfil }
\vskip 36pt Simulations  of some quantum 
gates, with  decoherence}
\author{V. Corato}
\author{P. Silvestrini}
\affiliation{Second University of Naples, 81031 Aversa, Italy}

\author{A. G\"orlich}
\affiliation{M. Smoluchowksi Institute of Physics, Jagellonian
University, Reymonta 4, 30-059 Cracow, Poland}

\author{P. Korcyl}
\affiliation{M. Smoluchowksi Institute of Physics, Jagellonian
University, Reymonta 4, 30-059 Cracow, Poland}
\author{L.~Stodolsky}
\affiliation{Max-Planck-Institut f\"ur Physik 
(Werner-Heisenberg-Institut),
F\"ohringer Ring 6, 80805 M\"unchen, Germany}

\author{J. Wosiek}
\affiliation{M. Smoluchowksi Institute of Physics, Jagellonian
University, Reymonta 4, 30-059 Cracow, Poland}

\date{Oct 2006}

\begin{abstract} Methods and  results for numerical simulations of
one and two interacting rf-\sq systems suitable for  \ad 
quantum  gates are presented. These are based on high accuracy
numerical solutions to the static and time dependent Schroedinger
equation for the full \sq \h in
one and two variables. Among the points
examined in the static analysis is the range of validity of the
effective two-state or ``spin 1/2'' picture. A range of parameters
is determined where the picture holds to good accuracy as the
energy levels undergo gate manipulations. Some general points are
presented concerning the relations between device parameters and 
``good'' quantum mechanical state spaces.

 The time dependent simulations allow the examination of suitable
conditions for \ad behavior, and permits the introduction of a
random noise to simulate the effects of \den. A formula is derived
and tested relating the random noise to the \de rate.
 Sensitivity to device and operating parameters for the logical
gates NOT and CNOT are examined, with particular attention to
values of the tunnel parameter $\beta$ slightly above one. It
appears that with  values of $\beta$ close to one, a quantum
CNOT gate is possible even with rather short \de times.

Many of the methods and results will apply to
 coupled  double-potential well systems in general.

\end{abstract}

\maketitle

\section{Introduction}
  In previous work we have described
quantum logic gates
based on the rf \sqn. The basic operation involved is an 
adiabatic
inversion, where the \sq reverses flux states under the sweep
of an external field   $\phi^{ext}$. This
 is equivalent
to the logical NOT  \cite{deco}.  When a
second  \sq is added whose flux can add or subtract from \px,
parameter ranges were found  for  the two
$\phi_1^{ext},\phi_2^{ext}$  so that  the two-\sq \sy   undergoes
a reshuffling of levels equivalent to the logical operation CNOT 
 \cite{ref1},\cite{pla}.

 In this paper we  present  a study of these
systems by  numerical simulations which enable us to examine these
processes in more quantitative detail. Among the points we can
examine is the validity of the ``spin 1/2 picture''.
 In the previous work we often
found good agreement with a
simplified ``spin
1/2 analogy"
 where the two lowest states of each \sq are treated as an
effective
two-state system. This  picture is very useful in
understanding and predicting  the behavior of the \syn s and  here
we
 examine its validity by   simulations
for the full  many-state  system.

A further question which can be studied in detail is \adyn.
This is the operating principle for our
quantum gates and it is necessary to know under which conditions it
holds. 

Finally, 
  we can use our programs to study the effects of \de
on the quantum gates. We will examine how to introduce \de as a 
noise signal and study its effects on our operations.

It should be stressed that our results would have been difficult 
 if not impossible to obtain without the extensive \sy of numerical
programs. Due to the sensitivity to the various parameters and the
subtleties of the tunneling problem, analytic methods would be
difficult and uncertain. With the high accuracy programs, a short 
run can replaces otherwise long,  complicated, and often
approximate formulas. 

\section{One variable System}
We begin by shortly reviewing our approach
~\cite{ref1},~\cite{pla} as applied to the one variable, one
 rf \sq \syn, at first without \den. This
will allow us to  fix the notation and parameters, and to
establish the connection  between the full \h
and  the ``spin 1/2 picture''.

\subsection{ Squid Hamiltonian } The \sq Hamiltonian in terms of
the
capacitance $C$ and inductance $L$ of the junction is ${\cal
H}={-1\over 2C
({\Phi_0\over2\pi})^2}{\partial
^2\over\partial \phi^2 }+U$ with $U=({\Phi_0\over 2\pi})^2 {1\over
L}\{{1\over
2}[(\phi-\phi^{ext})^2 ]+\beta cos\phi\}$, where $\beta=\frac{2\pi
LI_c}{\Phi_0}$, $I_c$ being the critical current for the
junction~\cite{rev}.
The variable $\phi$ is
the flux $\Phi$ in the \sq loop in flux quantum units 
$\phi=\Phi{2\pi\over\Phi_0}$, and furthermore shifted by $\pi$ so
that $\phi=0$ corresponds to the maximum at the  center of the
double potential
well $U$. Similarly $\phi^{ext}$ is an external  field,  where due
to the shift, $\phi^{ext}=0$ corresponds to a non-zero applied
field of size $\frac{\Phi_0}{2}$.
By factoring out an overall energy scale $E_0=1/\sqrt{ LC}$ so that
${\cal H}=E_0 H$  we obtain the dimensionless hamiltonian

\begin{equation}\label{hama1}
H={-1\over 2\mu }{\partial ^2\over\partial \phi^2}
 +V_0\{\tfrac{1}{2}[(\phi-\phi^{ext})^2 ]+\beta\,
cos\phi\}={-1\over
2\xi }{\partial ^2\over\partial \phi^2}
 +\xi \{\tfrac{1}{2}[(\phi-\phi^{ext})^2 ]+\beta\, cos\phi\}\;. 
\end{equation}
With  this  choice of the factor $E_0$ the ``mass'' $\mu$ and the
``potential'' $V_0$
are equal, hence the second form in terms of a common parameter
$\xi$, whose value is $\xi=\mu =V_0\approx 1030 \sqrt{{C/pF\over
L/pH}}$. In effect the capacitance C and inductance  L have  been 
exchanged for an energy scale $E_0$ and a dimensionless number 
$\xi$. We shall
discuss the physical meaning of $\xi$ below.

In the following we shall endeavor to express all energies in terms
of the general energy unit $E_0$ and all times in the time unit
$1/E_0$.
To converted these to dimensional units  one
multiplies by 
$ E_0=6.4\times10^{-4}eV\times (L/pH ~C/pF)^{-1/2}=1.0\times
10^{12}~
radians/sec~\times(L/pH ~C/pF)^{-1/2}= 7.7 K \times (L/pH
~C/pF)^{-1/2}$ and for the time
by
$1/E_0=1.0\times 10^{-12} seconds\times (L/pH
~C/pF)^{1/2}$ ~\cite{correct}. Thus results with the dimensionless
\hn, Eq~\ref{hama1}, which we will use in our computer simulations,
are to be converted to physical energies by multiplying by $E_0$. 
  Typical values $L=400 pH$ and $C=0.1pF$ for example, yield
$E_0=160\times 10^{9} radians/sec=1.2\,K$ and  $ \xi=16.3$, while
the time
unit is 
$6.3\times 10^{-12}~sec$. We will use these values for ``typical
examples'' or when absolute times or energies are needed. We shall
usually work with $\beta=1.19$,
which for $L=400 pH$ corresponds to a critical current of about
$1\, \mu A$.

To carry out   calculations with Eq~\ref{hama1} for the
eigenvalues and eigenfunctions, the numerical method employed 
a large basis of harmonic oscillator \wvs and  expanded the
cosine as a low order polynomial. Since in the harmonic oscillator
basis polynomials are simple, sparse, matrices, the problem is
reduced to 
algebraic manipulations and a matrix diagonalization. It was
usually
found that an expansion up to 8th order  for the cosine and a
basis of 256 oscillator
states gave numerical stability. A typical run   with a few hundred
basis vectors  on a Pentium
4  machine lasts less than a minute and resolves our smallest
energy splitting to better than two significant figures.

\subsection{Parameters of the Hamiltonian}
 Eq~\ref{hama1} contains three
parameters, \px, $\beta$ and $\xi$. The applied flux  \px ~
controls
 the asymmetry of the potential and for \px=0 the potential
is symmetric.   Fig.\,1, Left, shows an example with the
potential in this symmetric configuration. In this ``level
crossing'' situation the energy splitting of
the lowest pair of levels  is  just the tunneling energy
$\omega_{tunnel}$ and
so is quite small. Here it is 0.0044 and not clearly resolved on
the plot.
\begin{figure}
\begin{center}
\begin{tabular}{cc}
{{\includegraphics[width=0.5\hsize]
{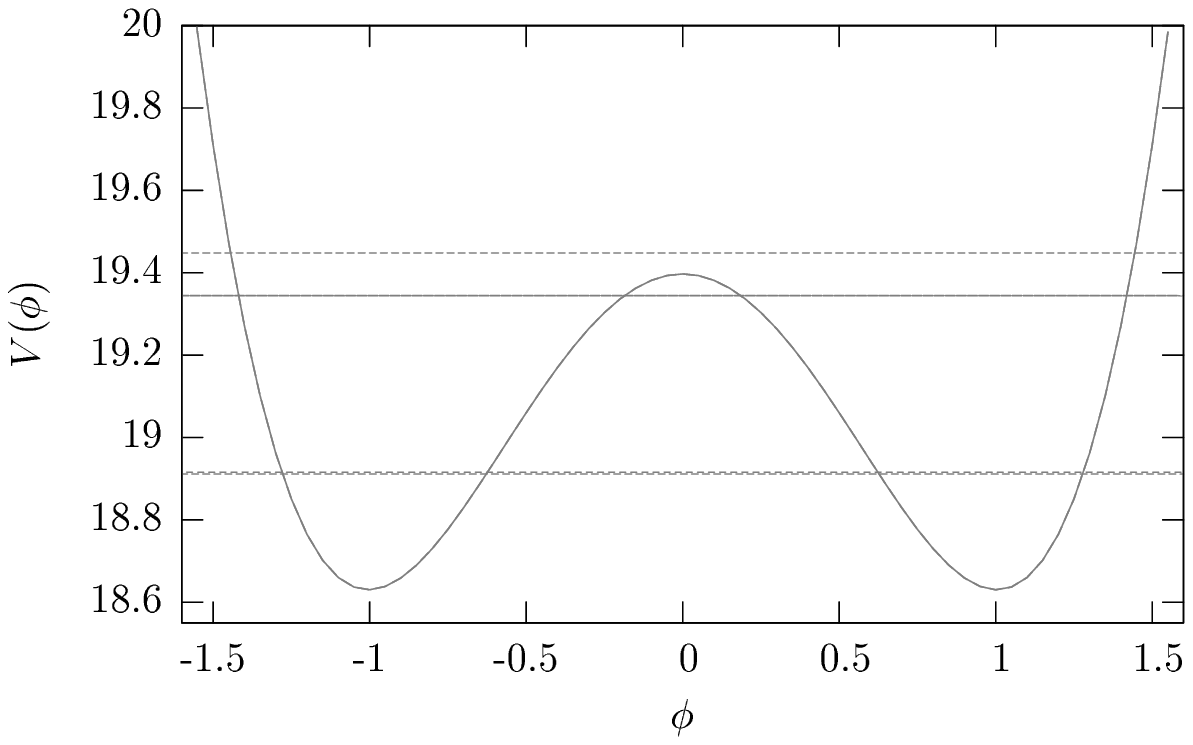}}}&{{\includegraphics[width=0.5\hsize]
{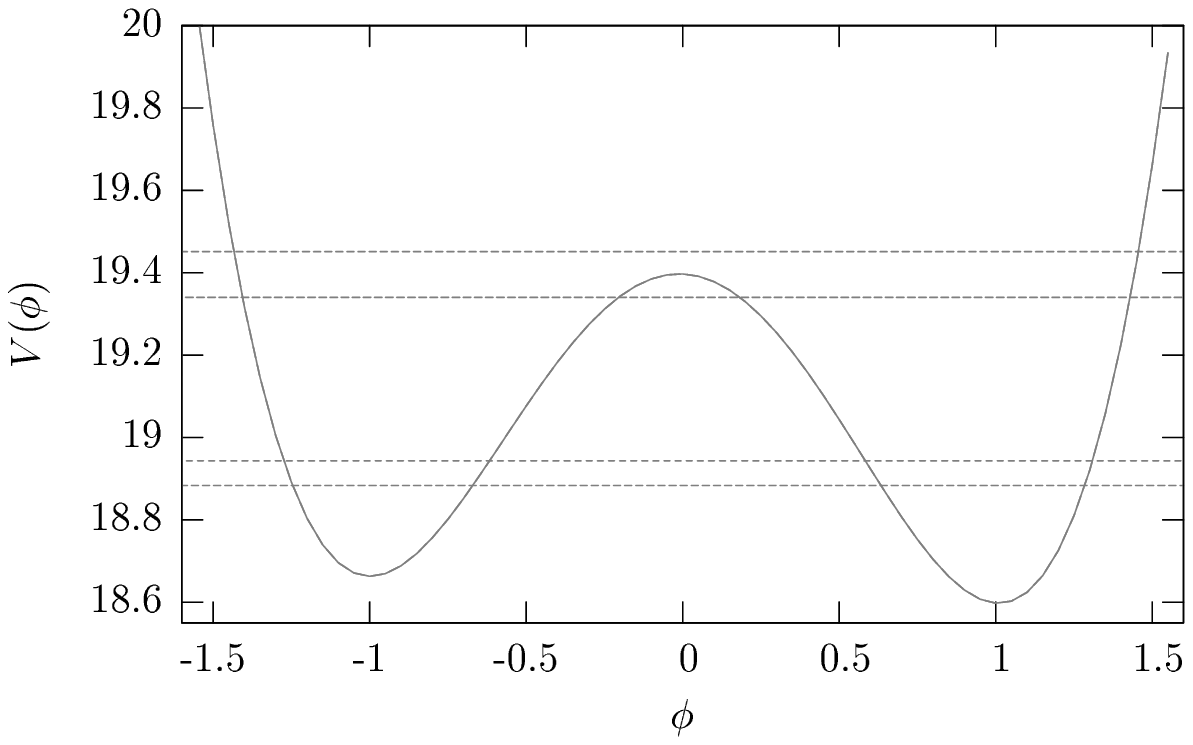}}}
\end{tabular}
\end{center}

\caption{Potential with first four energy levels according to
Eq~\ref{hama1}, for $\beta =1.19$
and $V_0=\mu=16.3$. Left:  The symmetric or ``level crossing''   
configuration with \px=0. 
The lowest line represents a pair of levels which are not clearly
resolved on the plot. Right: An asymmetric configuration with
\px=0.0020, showing a much greater pair separation. The horizontal
axis represents the reduced flux
$\phi$, the vertical axis the energy in units of
$6.4\times10^{-4}eV (L/pH ~C/pF)^{-1/2}$.}
\end{figure}

  Fig.\,1, Right,  shows, under the same
conditions, an asymmetric configuration with \px=0.0020.
To produce an adiabatic inversion or NOT,  
  \px ~begins at such a
non-zero value and  adiabatically  "sweeps'' to the opposite value.
The level splitting is now dominated by the shift of the potential
and not the tunneling. The splitting of the lowest 
pair is 0.060, an order of magnitude greater than in the symmetric
configuration.
  The \wvn s of the energy eigenstates are now concentrated in the
left or right well, while at ``level crossing'' 
they were $\pm$ linear combinations of these states. 

Turning to the parameter $\xi$,  raising its value both increases
the "mass" and the height of
the
potential  and so will generally lead to a greater
concentration of the \wv in one of the potential wells, as well as
a reduction of the tunneling. In the wells one has
roughly harmonic oscillator potentials, for which 
the ground state wavefunction is  $\psi\sim e^{- {1\over
2}\sqrt{c\mu V_0}
\phi'^2} $, where $\phi'$ means $\phi$ referred to its equilibrium
position and $c$ is a factor of order one. Thus $\xi$ gives the
spread of the wavefunction in the well; it controls 
how ``uncertain''
the flux is  when it is in a ``definite'' state. Similarly the
exponent in the  tunneling factor $e^{-\int dx \sqrt{2\mu
(V_0-E)}}$ increases 
with $\xi$ and reduces the tunneling. Therefore $\xi=\sqrt{\mu
V_0}$ may be viewed as a measure of
how "classical" the \wv
is, and 
 the relatively large value of $\xi$ we deal with indicates
that our component \wvn s are rather well defined and "classical"
in
this sense. That
is although we deal with linear combinations of states, these basis
states themselves are relatively localized, are rather
``classical''. Examples of this will be presented in Fig.\,9.

The parameter $\beta$, finally,  has a strong effect on the
tunneling since increasing its value 
  both widens and heightens the barrier. Thus increasing
 $\beta$ from 1.19 to 1.35 leads to
three pairs of well defined states below the barrier, with a tunnel
 splitting of only $2.0\times 10^{-6}$ for the lowest pair. In the
other direction, for $\beta$
somewhat less than one, there are no longer any states localized
around a
definite value of the flux at all. The most interesting region for
the present purposes appears to be the values of 
$\beta$ somewhat greater than one, and for most of our simulations
we shall use $\beta=1.19$.
With  $\beta =1.19, \xi=16.3$, the splitting at level crossing is
0.0044,
while the distance to the next set of levels is 0.48. This
difference of two orders of magnitude  provides  a factor of ten 
margin in manipulating the levels with still a factor of ten to the
next set. However in studying specific designs it may be of
interest to make a careful study of the behavior with respect to
$\beta$ in the vicinity of one, and we shall also briefly consider
$\beta=1.14$.

Finally, it should be observed that the energy splitting from our
lowest
 pair of states  to the next ones is on the order of the energy
unit,  which is 
$E_0\sim 1\,K$. Consequently  when working well below $ 1 K$, which
we shall
assume, one  
can expect only the lowest pair of states to be populated, and the
neglect of 
thermal over-the-barrier transitions to be justified.

\section{Identification with a Spin 1/2 System}

It is frequently a useful simplification to use the ``spin
picture'' where we treat   two closely separated levels, such as
the lowest pair in Fig.\,1, as the two
states of a  ``spin 1/2 system''~\cite{us}; and for the two qubit
\sy as two
such
spins. In this section we discuss the relation between this picture
and the full \hn.
The spin picture \h in the absence of noise or \de is 
\begin{equation}\label{haa}
H=\tfrac{1}{2}\,{\bf \sigma \cdot V}\;,
\end{equation}
where we drop an irrelevant additive constant.
 
We wish to use the two lowest states of Eq~\ref{hama1} as our
qubit, and to identify it with a \sy which can be effectively
described by the simple
Eq~\ref{haa}. In other words,
 we wish to  use the two lowest energy
eigenstates found from  the exact Eq~\ref{hama1},
with say  \px=0, to
span a two-dimensional   basis to set up the  spin picture.
We assume that as  \px ~ is varied over a small range  we stay in
this Hilbert space and
 always have to do with various linear combinations of the same
\wvn s. This 
 seems a plausible  assumption  when the tunneling energy is small
compared to
the
other level splittings, and we shall present  evidence for it
 below.

 Having made this assumption, the next question is which linear
combinations are to be used as the fixed basis--in the spin
language which linear combinations to chose as ``spin up'' and
``spin down'' along an abstract ``z axis''.   As  for the
flavor with neutrinos and K mesons, or the handedness with chiral
molecules~\cite{us}, we wish to chose these basis states to be
eigenstates of a definite, externally measurable, property of the
\syn.
Here we choose the direction of the flux, i.e. the  current  in the
\sqn ,  corresponding to the 
\sy localized either in the left or the right potential well.
 Due to the tunneling these states are not energy eigenstates and
so not stationary; they  will generally
undergo  oscillations in time.

This choice  implies that  the basis states, those to be 
identified with ``spin up'' and ``spin down'' along the ``z axis'',
are chosen to be eigenstates of the flux $\phi$. In terms of the
full   Eq~\ref{hama1}, these are \wvn s   concentrated in one 
potential well only.  Now in our two-dimensional space spanned by
the two lowest eigenstates, the ``position'' variable
 $\phi$ is a $2 \times 2$
matrix. This matrix  can be expanded in terms of the Pauli
matrices. Our choice that the basis states are eigenstates of
$\phi$ means that the abstract axes are defined such that $\phi$ is
proportional to the diagonal Pauli matrix $\sigma_z $
  
\begin{equation}\label{axa}
\phi\sim \sigma_z\;.
\end{equation}

Furthermore,
we can  approximately establish the constant in this relation by
using the fact  that for the so-defined eigenstates of $\sigma_z$
the  \wv
is approximately localized.  We then approximately
know the value of $\phi$, since operating on one of these states,
for example $\vert R >$ for a state concentrated in the right well,
we expect $\phi\vert R > \approx\phi_c\vert R >$
where the number $\phi_c$ is the value of the variable $\phi$ where
the wavefunction is concentrated, say has its maximum value.
 Since $\sigma_z$ operating on one of the localized states has been
defined as $\pm 1$, Eq~\ref{axa} becomes
\begin{equation}\label{ax}
\phi\approx \phi_c \sigma_z \, 
\end{equation}
As may be seen in Fig.\,1, $\phi_c$ will usually be
roughly one.

With this identification we  now proceed to analyze the full \h
Eq~\ref{hama1}.
Since we  always take $\phi^{ext}$ small, it is a good 
approximation to write Eq~\ref{hama1} as
\begin{equation}\label{hama2}
H\approx {-1\over 2\xi }{\partial ^2\over\partial \phi^2}
 +\xi\{\tfrac{1}{2}\phi^2 +\beta\, cos\phi\}-\xi\phi^{ext}\phi\; .
\end{equation}
Eq~\ref{hama2}  represents the total Hamiltonian as the \h for the
case of the symmetric or ``level crossing'' potential where \px =0
and an asymmetric term
proportional to $\phi$. Evidently, in view of Eq~\ref{ax} and the
symmetry of the first part of Eq~\ref{hama2},   the last term in 
Eq~\ref{hama2} is to be identified with the $\frac{1}{2}
V_z\sigma_z$ of
the spin \h Eq~\ref{haa}.

From this observation we can, by using Eq~\ref{ax}, identify the 
value of $V_z$  in Eq~\ref{haa} with the parameters of the full \h 
Eq~\ref{hama1} as

\begin{equation}\label{est}
V_z \approx 2 \xi \phi^{ext} \phi_c \; .
\end{equation}
With $\phi_c$ close to one, we anticipate $V_z \approx 2 \xi
\phi^{ext}$. Below we show a more
exact calculation of $V_z$ from the
numerical results with the full \hn.

The quantity from the  full \h to be associated with the
magnitude  V of $\bf V$ in the  spin picture is easy to identify.
 The eigenvalue splitting
of $\tfrac{1}{2}{\bf \sigma \cdot V}$ is always V. Thus  given a
numerical calculation with Eq~\ref{hama1} yielding the level
splittings, we anticipate
\begin{equation}\label{spl}
{splitting}=V=\sqrt{V_z^2+V_x^2}\;,
\end{equation}
where $splitting$ is the energy difference of the two lowest
states.
(A $V_y$ does not enter into our considerations since all 
quantities are real and $\sigma_y$ would involve imaginary
quantities.)

\begin{center}
\begin{table}

\begin{tabular}{|l|l|l|l|l|l|l|l|} 
\hline
$\beta$&$\xi$~~~& $\phi^{ext}$ & Splitting
&$sin\theta_{V}$&$\theta_V(rad)$&$
\theta_{\phi}$(rad)&Completeness\\
\hline
\hline
1.19&16.3&0.0 &0.0044&1.0&1.6&1.6&1.0\\
\hline
~&~& 0.000030&0.0045&0.98&1.4&1.4&1.0\\
\hline
&&0.00010&0.0053&0.83&0.97&0.97&1.0\\
\hline
&&0.00030&0.010&0.44&0.46&0.46&1.0\\
\hline
&&0.00070 &0.021&0.21&0.21&0.21&1.0\\
\hline
&&0.0020 &0.060&0.073&0.073&0.075&1.0\\
\hline
&&0.0070 &0.21&0.021&0.021&0.027&0.99\\
\hline
&&0.010 &0.30&0.015&0.015&0.024&0.99\\
\hline
&&0.015 &0.45&0.0099&0.0099&0.042&0.95\\
\hline
&&0.019 &0.54&0.0081&0.0081&0.59&0.23\\
\hline
\end{tabular}
\caption{ Quantities from the exact numerical simulation for
comparison with those of the ``spin 1/2 picture''. The energy
splittings are to be seen in relation to the distance to the next
set of
levels, which is about 0.6 units. The angle
$\theta_{V}$ refers to the angle the $\bf V$ vector makes with the
z-axis, while the angle
$\theta_{\phi}$ refers to the same angle inferred from the spinor
eigenfunctions, as described in the text. ``Completeness''
characterizes
how well the two states in question remain in the same state space
as \px is varied. In general, the results indicate that the two
state picture holds
 to about \px $\approx 0.01$. Blank spaces in tables imply
repetition of
previous values.}
\end{table}
\end{center}

The x-component $V_x$ represents the tunneling energy 
$\omega_{tunnel}$, or the splitting resulting from the first,
symmetric
part of Eq~\ref{hama2}, 
-- the \h at ``level crossing''. 
 We can therefore obtain $V_x$ from 
the numerical evaluation at \px$\propto V_z=0$: 

\begin{equation}\label{splt}
V_x=\omega_{tunnel}={splitting}\vert_{(\phi^{ext}=0)} \;.
\end{equation}  
The numerical value of  $V_z$  then follows   from those for $V$ 
and $V_x$. 

\subsection{ Energy level behavior }
We can test this picture, where $V_x$ and $V_z$ are  treated
as approximately independent quantities, by plotting the splittings
from the full numerical calculation  vs \px\, to check if they have
the expected form $\sqrt{V_z^2+V_x^2}$, with $V_z$
proportional to \px\, and $V_x$ a constant. This is shown in
Fig.\,2. There is an
excellent fit to this form with the fit constant for $V_x$ equal to
the value at level crossing. In addition the fit coefficient C in 
$V_z=$2C\px\, is 14.9,
while  from the estimate Eq~\ref{est} we would
expect $C\approx V_0\phi_c$ which  with $\phi_c\approx 1$ is
16.3. Or if we adjust $\phi_c$ to make the identification exact,
we need $\phi_c=0.91$. Inspection  of a plot of the \wv shows that
its maximum is indeed close to this, at about $\phi=0.88$.

  Table 1 shows some  of these values and also those for some
larger \px .   Deviations from
the linear behavior $ splitting \propto $\px~
begin to set in at about \px = 0.015, where the
splitting
is near a half unit. As would be expected, this  is on the order of
the distance to the next set of
levels, as seen in Fig.\,1. Thus, for the behavior of the
energy levels at small \px \,, there is quite good
agreement between the
numerical calculations with the full \h and the ``spin 1/2
picture''. Experimentally, a plot equivalent to Fig.\,2 has been
mapped out to about \px=0.008 \cite{mooj1}.

\subsection{ Rotation angle}
As another  check on the simple two-state picture,
 we can consider two independent ways of finding
the angle the ``spin''  makes with the abstract z axis. One way is
to use
$V_x$
and the magnitude $V$ from the energy splittings. These are given
by  Eqs~\ref{splt} and \ref{spl} and the resulting $\theta_V$ from
$sin\theta_V=V_x/V$ is given in  Table I.   A second way,
using only
the \h at a given \px , is to find the angle of rotation from the
energy eigenstates to the ``$\sigma_z$'' eigenstates. If the angle
that
$\bf V$ makes with the z axis is $\theta$, then the eigenstates
$v_{\pm}$ of the spin \h  are given in terms of the spin-up, spin
-down
states $u_{\pm}$ as $(cos\frac{\theta}{2}u_+
+sin\frac{\theta}{2}u_-)
$ and $(-sin\frac{\theta}{2}u_+ +cos\frac{\theta}{2}u_-)$. Thus by
finding the rotation from  the $v$ to the $u$ we may determine
$\theta$. Numerically, this is done by computing the 2x2 matrix of
$\phi$ in the energy eigenstate basis and finding the rotation
necessary to diagonalize it. The results are shown in
the Table as $\theta_{\phi}$. Disagreement between the two
methods begins to appear around \px =0.01, where the splitting is
0.3 energy units. This comparison is perhaps 
more sensitive than
that using the energy levels alone. However the disagreement is
only significant when the angle is small.

\begin{figure}
{{\includegraphics[width=0.5\hsize]
{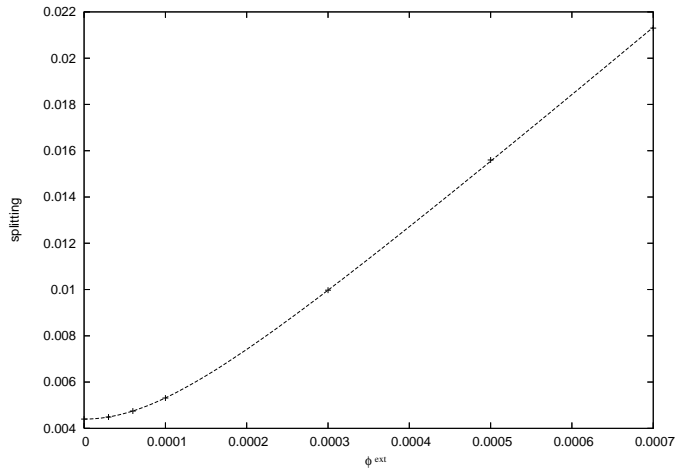}}}
\caption{Level splitting as a function of \px \,  for small
 \px ,   from
the numerical
calculation with the full \hn.
A good  fit  with the ``spin 1/2'' form
$splitting=\sqrt{(V_z^2
+( V_x)^2}~=\sqrt{(2C\phi^{ext})^2
+( V_x)^2}$ is obtained, yielding   $V_x=0.0040$ and $C=14.9$. C is
in agreement with the prediction
Eq~\ref{est}.}
\end{figure}

\subsection{Hilbert space completeness}
Finally, as another check, we can try to directly see if the \sy
remains in
the same two dimensional Hilbert
space as \px\,  is varied. We examine this by
comparing the spaces spanned by the  two lowest eigenstates of the
\hn s with
different \px. The worst overlap between the eigenstates of the \h
with \px =0 and those with the  non-zero \px  are listed in the
column ``Completeness''. The method  will be
explained below   when the two dimensional  case is discussed. 
  Significant
deviations again appear around \px=0.015  and one notes  a sudden
change as mixing with the next set of principal states
 becomes important. 

   In summary, the two state picture seems to work well with  the
parameters  used here, up to 
 about 0.01 for \px, or pair
splitting 
$\sim 0.2$ energy units, to be compared with the principal level
spitting of $\sim 0.6$
units. Inside this range, the picture that we always deal with
different linear combinations of the same two states while the \h
varies seems to be justified.

\section{ Adiabaticity}
 
Our gates operate by a sweep of the externally applied  \px.  The
speed of a sweep is of course relevant to how fast
a device or a set of devices might operate. Probably more important
than the simple speed,
however, is its connection with the \de question. The \de is
characterized by a rate (our D below). Therefore  fast gates,
giving  the \de less time to act, are favorable from the point of
view of \den.

 Although in principle very fast sweeps thus seem desirable, this
is not possible without violating the adiabatic
condition  upon which the gate operations are based.
 It is therefore
 of interest to find out how fast  sweeps  can be performed
without violating adiabaticity. 
Using the simulation  programs we can study this point
 in detail.
First we  examine the simple case of the
adiabatic inversion or  NOT, using one \sq without \den.

To deal with the time dependent \hn s numerically, repeated
iterations of $(1+iH \Delta t)$  were employed, applied to \wvs
found by the
methods of the static calculation described above.
To save time in such runs the matrix H was calculated not in the
original large
oscillator basis  but in a reduced basis of the few lowest energy
eigenstates. The results
could be checked by incorporating more states into this ``second
cut'' basis. Usually, as might be expected from the arguments
around Table I, two states were sufficient in one variable and four
states in the two variable problem.

According to the estimate given in ref~\cite{ref1},
 adiabaticity is guaranteed when the sweep time $t_{sweep}$ is
sufficiently long such that 
\begin{equation}\label{adbt} {\epsilon \over t_{sweep}} <<
\omega_{tunnel}^2\,,
\end{equation}
where $\epsilon$ is the initial energy level splitting (same
as the  initial difference of the minima of the
potential wells  to about 10\%).
 The condition is
sensitive to
$\omega_{tunnel}$ because the violation of \ady  takes
place essentially at  level crossing. Violation  occurs when the
rate of change
is too fast; and the left-hand-side of Eq~\ref{adbt} characterizes
this velocity of change of
the \syn.  Note we cannot
make the left-hand-side  smaller by simply reducing $\epsilon$. If
we want to retain a good definition of the original
state as approximately a state of definite flux, one
requires    $V_z/V_x>>1$ or  $\epsilon>>\omega_{tunnel}$, that is 
a substantial initial splitting compared to the splitting at
``crossing''.

\begin{figure}[h]
{{\includegraphics[width=0.5\hsize]
{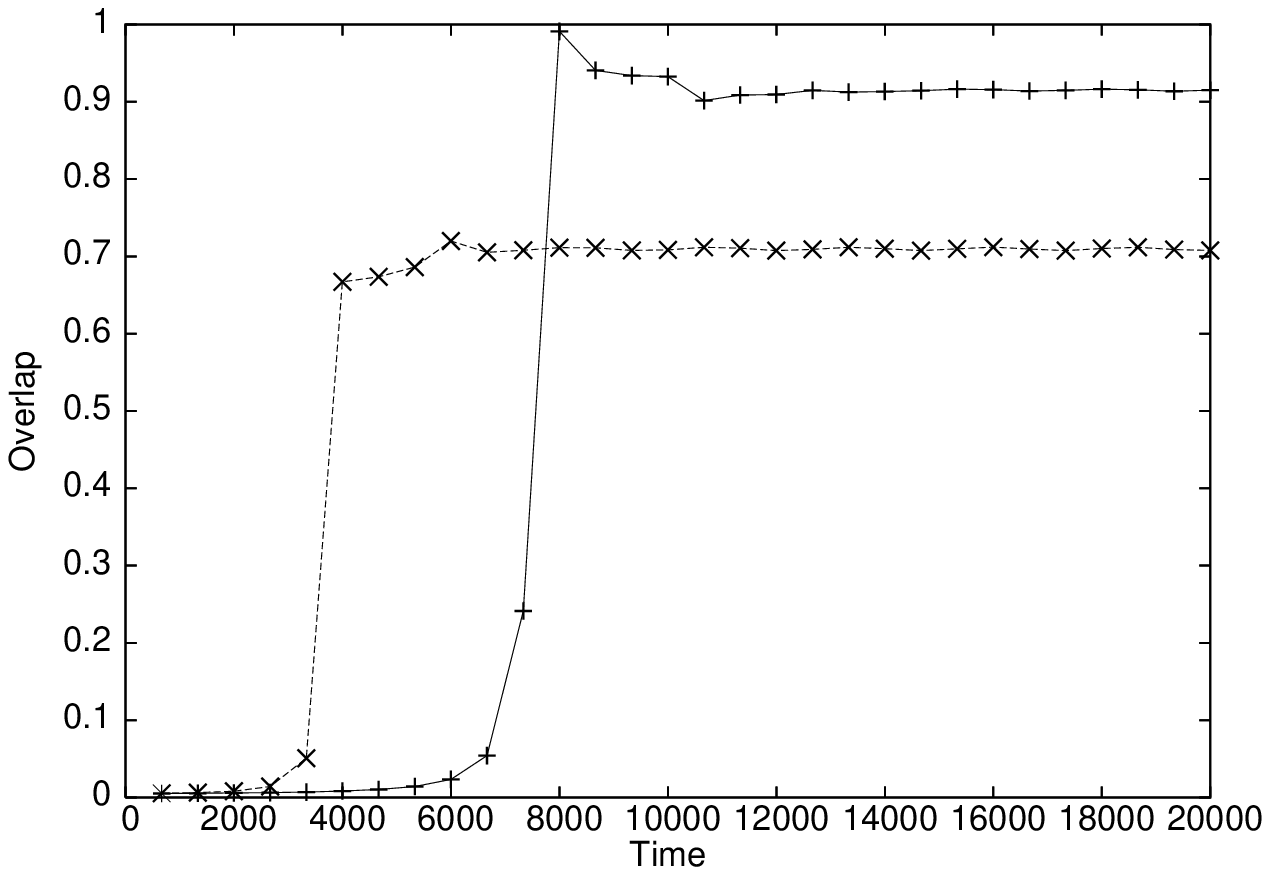}}}
\caption{Study of adiabaticity. The behavior of the scalar product
(squared) between an
evolved \wv  and the  target  eigenstate
of  the final \hn. The 
evolved \wv is calculated from the Schroedinger equation with the
\h Eq~\ref{hama1} containing a time-varying \px\, 
 that  reverses sign.  Two sweeps are shown,
one (lower) with
$t_{sweep}=7500$ time units and one (upper) with $t_{sweep}=15000$
units.  ``Level crossing''  or \px=0 occurs
at $\frac{1}{2} t_{sweep}$. The \ad
inversion or reversal
of the flux state of the \sq is  more ``successful'' for
the slower
sweep. Conditions as for the $\beta=1.19$ case in Table II. }
\end{figure}

We may characterize the ``success'' of an \ad inversion by starting
with a \wv which is an eigenstate of the initial potential,
evolving it in time with the changing potential, and finally
comparing it 
with the stationary  \wv to which it should arrive, namely the
eigensolution for the final
potential. In the case of  perfect \ady the two \wvn s  should be
the same. Thus we can gauge the loss of \ady  by the deviation from
one of
 the scalar product of the evolved \wv  with the ``target'' \wvn .
 Fig.\,3  shows the evolution of this ``overlap'', in terms of  the
scalar product squared. A relatively
fast  ($t_{sweep}=7500$ time units)  and a  slower sweep
($t_{sweep}=15000$) are shown. 
 While the final overlap is about 0.9
for the slower sweep, it only reaches 0.7 for the faster sweep. 
Table II shows this final overlap for a choice of parameters.

 From 
 Eq~\ref{adbt} one can form the \ady  parameter $\omega_{tunnel}^2
t_{sweep}/\epsilon$, which is shown in the next-to-last column of
the Table.
This parameter, which compares the splitting at ``crossing'' with
the ``velocity''
${\epsilon / t_{sweep}}$ of
passing through the crossing,  will be recognized as that which
arises in the Landau-Zener theory \cite{lla}. As would be expected,
 one observes that  values of the parameter of order  one
characterize the transition from \ady to non-\adyn, and
 that similar values of the parameter give similar results.
However, there appear to be some small deviations, (see the 0.90
values),
showing the usefulness of detailed numerical calculations.
  A value of about 4 or greater  for the \ady 
parameter appears necessary to
achieve a 90\% success probability, while a value near one gives
the 50\% point. 
With the typical time unit of  $ 6.3\times 10^{-12}s$  a sweep of
thousands of time units  corresponds to some nanoseconds. 
 Table II  exhibits the great sensitivity of
$\omega_{tunnel}$  to $\beta$, a reflection of the exponential
nature of the
tunnel splitting.

In the simulations the sweep is carried out by simply letting \px
~vary at a constant rate as it goes from some initial value to the
same  value with the opposite sign; that is, the sweep is simply
linear with
a constant ``velocity''. 
 In more refined versions of the sweep
procedure, it is
conceivable that  special waveforms  could be used so
as to pass through the dangerous vicinity of \px=0 slowly while the
overall sweep is very fast. However, it should be kept in mind that
the \de is also most effective in the vicinity of ``level
crossing'' (see below). All calculations presented here are
performed with
simple linear sweeps.
\begin{center}
\vskip.5cm
\begin{table}
\begin{tabular}{|l|l|l|l|l|l|l|l|l|} 
\hline
$\beta$&$\xi$& initial \px&
$\epsilon$~&$\omega_{tunnel}$& $t_{sweep}$&$\omega_{tunnel}^2
t_{sweep}\epsilon^{-1}$&final~overlap\\
\hline
\hline
1.14&16.3&0.002 &0.055&0.027&1 000&13&1.0\\
\hline
1.14&& &0.055&0.027&300&4.0&0.90\\
\hline
1.18&& &0.058&0.0065&10 000&7.3&0.98\\
\hline
1.18&& &0.058&0.0065&4 000&2.9&0.78\\
\hline
1.19&& &0.060&0.0044&15 000&4.0&0.90\\
\hline
1.19 &&&0.060&0.0044&7 500&2.4&0.70\\
\hline
1.19 &&&0.060&0.0044&4 200&1.4&0.50\\
\hline
1.21&& &0.063&0.0019&80 000&4.6&0.90\\
\hline
1.21&& &0.063&0.0019&40 000&2.3&0.68\\
\hline
1.29&& &0.075&0.000045&120 000&0.0032&0.016\\
\hline
\end{tabular}
\caption{Tunnel splitting $\omega_{tunnel}$ as a function of
$\beta$ and its effect
on \adyn.  One observes a rapid change in $\omega_{tunnel}$ with
$\beta$. The  resulting effect on the ``success'' of an \ad
inversion is measured
by  the ``final overlap'' as in Fig.\,3. Similar values of the \ady
parameter give similar results for the overlap.
 With our ``typical parameters''  one thousand time units is 6.3
ns.}
\end{table}
\end{center}

\section{Decoherence}

We shall  present a
simple model for the \de suitable  for numerical simulation and
then examine
its effects in various contexts. We first apply the model in the
``spin 1/2 picture'' where it can  be handled analytically and then
compare with numerical calculations with the full \hn.

\subsection{Introduction}

To discuss decoherence it is necessary to introduce the density
matrix $\rho$ \,\cite{ll}. For our present purposes we
think of $\rho$  as a matrix arising
from an average
over different wavefunctions:

\begin{equation} \label{av}
\rho=\overline{\psi \psi^{\dagger}}={1\over N}\sum_{a=1}^N {\psi^a
\psi^{a\dagger}}\; ,
\end{equation}
where each $\psi^a$ is a column vector and so $\rho$ is a matrix.
Our model will  simulate the \de as a  noise in the
Hamiltonian,
producing different wavefunctions in the evolution from an initial
time to a final time. We then average over these wavefunctions
according to  Eq~\ref{av} to obtain the density matrix.

We recall the description of \de for the two-state system 
 when thermal over-the-barrier transitions
are
neglected \cite{us}.  The \de is characterized by the
parameter $D$  arising in
the effective
``Bloch equation''  giving the evolution of the \dmn .
That is, the 2x2 density matrix is written in terms of
the Pauli matrices $\sigma$ 
 
\begin{equation}\label{rhomata}
\rho= \tfrac{1}{2} (1+{\bf P}\cdot {\bf \sigma}) \; ,
\end{equation}
where the information about the state of the system is in the
``polarization vector'' $\bf P$,  the average value of the
abstract ``spin'', ${\bf P}= Tr[\rho {\bf \sigma}]$.

The time dependence of $P$ is given by the equation
\begin{equation}\label{pdota}
\dot {\bf P}={\bf P}\times {\bf V}- D\,{\bf P}_{T}\; .
\end{equation}
   where ``T'' stands for ``transverse'',  and
means the components of ${\bf P}$ perpendicular to the direction in
the abstract space chosen by the external perturbations. In the
following we  take the outside
perturbations to be along the   abstract z-axis associated with the
basis states defined earlier. $\bf P_T$  then    refers to the x,y
components and
represents the degree of quantum phase coherence between the two
basis states $u_{\pm}$.
 The
shrinking of ${\bf P}_{T}$ induced by $D$ in Eq~\ref{pdota}
signifies a loss
of phase coherence between the basis states.
 The parameter D thus characterizes the \de rate of the \syn. We
shall also use $t_{dec}=1/D$ to refer to the \de time.

An important conclusion one can draw from
 Eq~\ref{pdota} is that the major contribution to the \de occurs at
``level crossing''. In a sweep passing through a ``crossing'' the
$\bf P$
vector swings from ``up'' to ``down'' so that at ``crossing'' it is
purely  horizontal, with a  large $\bf P_T$. If we take  
the scalar product with
$\bf P$ in Eq~\ref{pdota} 
\begin{equation}\label{pdotb}
\tfrac{1}{2}\frac{d}{dt}{\bf P}^2={\bf P}\cdot \dot{\bf P}=-
D\,{\bf P}\cdot{\bf P}_{T}=-D {\bf P_T}^2\; .
\end{equation}
Since the departure of  $|\bf P|$ from 1 measures the loss of 
coherence,  the equation  shows that the
greatest ``shrinkage'', i.e.   loss of coherence,
  occurs when $\bf P$ is transverse, at level crossing.

\subsection{Random field  in the full \h}

 To simulate the \de numerically we adopt a
random field approach, where we suppose  a small random
time-dependent perturbation 
present in the \hn. This gives different realizations $a$ of the
\h  $H^a$. Starting with a given initial \wvn, we evolve
it with $H^a(t)$ to yield a final $\psi^a$.   Multiple repetitions
of this procedure, with an
average over the different realizations $\psi^a$  according to Eq
\ref{av}, gives
the density matrix originating from the initial pure state.
This procedure can be implemented on the
computer in a straightforward way.

 We shall model the
perturbations due the external environment  as
a kind of flux noise, with a random noise  $\cal N$ added to \px
~such that
\begin{equation}\label{noi}
\phi^{ext} \to \phi^{ext}+{\cal N}^a(t)\;.
\end{equation}

The external flux is of course not the only possible source of
noise, in 
principle one may envision fluctuations of any of the parameters
in the \hn.
 However the flux noise
  may well  be a good representation of the \de and may also
correspond to the
 main source of external noise in 
actual experiments. Naturally,   many of our general conclusions
 will remain valid regardless of the specific  origin  of the
noise/\de. 

Neglecting  ${\cal N}^2$ effects, Eq~\ref{noi}  implies
$\tfrac{1}{2} (\phi-\phi^{ext})^2 \to 
\tfrac{1}{2}(\phi-\phi^{ext})^2
-\phi {\cal N}^a(t)$,
so that the \h becomes
 
\begin{equation}\label{hama3}
H^a={-1\over 2\mu }{\partial ^2\over\partial \phi^2}
 +V_0\{\tfrac{1}{2}[(\phi-\phi^{ext})^2 ]+\beta\, cos\phi\} -V_0
\phi{\cal
N}^a(t)\; .
\end{equation}

 To interpret the additional term we
use the
identification
Eq~\ref{ax},  relating  $\phi$ to the $\sigma_z$ of the spin
picture:

\begin{equation}\label{bid}
-V_0
\phi{\cal
N}^a(t)\approx V_0 \phi_c{\cal N}^a(t) \sigma_z =\sigma_zB^a(t)\;,
\end{equation}
where we call the quantity $-V_0 \phi_c{\cal N}$ the random field
$B$. 
Thus in the spin picture the noise term has the interpretation of
an additional random ``magnetic field'' $B$ applied to the z-
component of the spin. We take this random field to have average
value $\overline B=0 $.

\subsection{Random field in the spin picture}

We begin with  an analysis within the spin 1/2 picture, which can
be handled
analytically and which provides an orientation for the full 
simulation. The \h of the spin picture is now

\begin{equation}\label{ha}
H^a=\tfrac{1}{2}{\bf \sigma \cdot V} + B^a\sigma_z\;.
\end{equation}

We first try to
 establish the connection between the \de parameter $D$ in
Eq~\ref{pdota} and the statistical
properties of $B$. 
Writing a wavefunction in terms of  ``up''
and ``down'' basis states $u_{\pm}$  as $\psi=\alpha u_++\beta
u_-$,  the contribution to the \dm from one instance in Eq~\ref{av}
is
\begin{equation}\label{rhom}
\rho=\pmatrix{ \alpha \alpha^* & \alpha\beta^*\cr\alpha^*\beta &
\beta\beta^*}\; ,
\end{equation}
where $\alpha \alpha^* +\beta\beta^*=1$ from  normalization.
Taking the average and comparing  with Eq~\ref{rhomata}, one has 

\begin{equation}\label{ps}
P_x={\rm Re}~ \overline {\alpha\beta^*}~~~~~~~~~~~~~P_y={\rm Im}~
\overline {\alpha\beta^*}
\end{equation}
 We now assume that the frequencies associated with the noise are
much above those connected with the slow coherent rotations induced
by
$\bf V$, so that $\bf V$ maybe thought of as effectively "turned
off" for
the calculation of $D$.
 In this case the solution of the Schroedinger equation is simple.
If  at $t=0$ we have the state $\psi=\alpha_0 u_++\beta_0 u_-$,
then
at time t 
\begin{equation}\label{psit} 
\psi^a(t)=\alpha_0~e^{i\int_0^t B^a dt} u_++\beta_0~ e^{-i\int_0^t
B^a dt} u_-\;,
\end{equation} 
Putting $\alpha_0$ and $\beta_0$ real  so $\bf P_T$ is initially
along the x-axis one has
according to Eq \ref{ps}

\begin{equation}\label{pxx}
P_x= (\alpha_0\beta_0){\rm Re}~ \overline {~e^{2i\int_0^t B dt}~}
\end{equation}
Assuming  the random field is of small amplitude permits the
expansion
\begin{equation}\label{pxa}
P_x= (\alpha_0\beta_0) {\rm Re}~ \overline {e^{2i\int_0^t B
dt}}\approx  (\alpha_0\beta_0)(1- 2~ \overline {({\int_0^t
B dt})^2})= (\alpha_0\beta_0)( 1-2\int_0^t\int_0^t dt'' dt'
\overline { B(t'') B(t')}~)\; ,
\end{equation}
where the linear term  vanishes with $\overline B=0 $.
We  thus have to do with the autocorrelation function
$ \overline { B(t'') B(t')}={1\over N}\sum_a { B^a
(t'') B^a(t')}$
where $B^a$ is a particular realization and $N$ the number of
 realizations. 

 As is known in the theory of random noise \cite{lu} a
correlation function of this type under stationary conditions
behaves such that
$\int_0^t\int_0^t dt'' dt'\overline { B(t'') B(t')}=t
\int_{-\infty}^{\infty} \overline { B(t)
B(0)}dt=2\,t\int_{0}^{\infty}
\overline { B(t) B(0)}dt $.

Now with $V=0$ and using, as follows  from Eq~\ref{pdota}, that
$P_x=P_{x 0}e^{-
Dt}\approx  (\alpha_0\beta_0)(1-Dt)$, comparison with Eq~\ref{pxa}
yields the identification
\begin{equation}\label{id}
 D= {1\over t}2\int_0^t\int_0^t dt'' dt'\overline { B(t'') B(t')}
=4\int_0^{\infty}\overline { B(t) B(0)} dt\;. 
\end{equation}
Thus in the ``spin 1/2 picture'' $D$
is given by the integral of the autocorrelation function of the
random field. Or,  regarding  $\sigma_z  B^a$ as a  perturbing
energy  $\delta E$, one can
also say $D=4\int_0^{\infty}\overline { \delta E(t) \delta E(0)}
dt$.

 As is evident from this short derivation, or from the assumptions
used in the original derivation \cite{us} of Eq~\ref{pdota}, it is
assumed that the frequencies of the random perturbations entering
into $D$ are high compared to the slow coherent motions induced by
$\bf V$. In the thermal context, where one anticipates the
perturbing frequencies to be on the order of the temperature, this
means that the energy splittings induced by $\bf V$ are assumed
small compared  (k=1 units)  to  the temperature. For this reason
it is consistent that in Eq~\ref{pdota} the density  matrix relaxes
to the identity and not to the form that would be given by a
Boltzmann factor.
Similarly, low frequency instrumental noise in the laboratory would
be better treated as an additional contribution
to $\bf V$ rather than being incorporated into $D$. In our
simulations we  always treat the noise as being of high frequency
in this sense.

\subsection{Modeling  of the noise}
A simple  noise model suitable for numerical simulation is

\begin{equation}\label{bc}
 {\cal N}(t)=\eta (t) \Delta\;,
\end{equation}
where $\eta$ is a random sign $\eta=\pm 1$, and $\Delta$ a positive
magnitude. Let $\delta t$ be a certain small time interval during
which
$\eta$ is constant and  let the probability that there is a sign
switch for the next interval be $p_{sw}$ (and to remain unchanged
$1-p_{sw}$). This procedure, in the limit of small  $p_{sw}$ and
$\delta t$, leads to an
exponential distribution for the noise pulse lengths 
and an autocorrelation function
\begin{equation}\label{auto}
\overline{\eta(t)\eta(0)}=e^{-2 p_{sw}t/\delta t }=e^{-2\alpha t }
\end{equation}
where $\alpha= p_{sw}/\delta t$.
  Introducing the noise power to characterize the frequency
content of the noise signal \cite{lu}, such an autocorrelation
function leads to the noise power spectrum
$\int_0^{\infty} dt~ cos\omega
t~\overline{\eta(t)\eta(0)}={\omega_c\over\omega_c^2 +\omega^2} $,
   with $\omega_c=2\alpha=2 p_{sw}/\delta t$ a cutoff frequency.
This spectrum is  roughly constant  (``white noise'') up to
the cutoff at about $\omega_c$ and then
falls off as $\omega^2$ at higher frequencies.
We attain the highest  cutoff, the closest approximation to
infinite frequency white noise, by  choosing the switching time
$\delta t$
as small as possible i.e., equal to the program step
and with the switch probability  $p_{sw}=1/2$. In this case 
 $\omega_c=1/program~step$.

For Eq~\ref{id} we need the time integral of the autocorrelation
function, which is
\begin{equation}\label{nato}
\int_0^{\infty}\overline{{\cal N}(t){\cal
N}(0)} dt= \Delta^2/\omega_c \; ,
\end{equation}

Using Eq~\ref{id} and recalling the definition of $B$ in 
Eq~\ref{bid},
one obtains,  finally
\begin{equation}\label{d21a}
 D=4 (V_0 \phi_c)^2 {\Delta^2\over \omega_c}\; .
\end{equation}
The $1/\omega_c$ behavior originates in the fact that one is 
essentially performing a random walk in the phase of the \wvn, and
the step
length of this random walk is $\sim \Delta\, \delta t \sim \Delta
/\omega_c$.
In the next section we check this prediction from the spin 1/2
picture against simulations with the full \hn.

\subsection{Tests of the noise simulation with the full \h}

  We now 
proceed   numerically, using the \sq \h Eq~\ref{hama3} . An
initial state is evolved with
Eq~\ref{hama3} 
with a given noise
realization  ${\cal N}^a$. Repeating this procedure
 many times  a final density matrix is obtained by
averaging over different realizations
according to Eq~\ref{av}. If the two-state approximation is good,
we should observe a \de rate in agreement with Eq~\ref{d21a}.

The numerical calculations were again carried out by   repeated
iterations of $(1+iH \Delta t)$. As few as
30 samples of the random field often give good (10\%) results.
A time step $\Delta t\approx 1$ was usually found sufficient.  For 
higher accuracy more samples and smaller time steps can be used.

It should be noted   that the \dm arising here from 
Eq~\ref{av} has a slightly different significance than    in
the ``spin 1/2 picture''.   Since we now work with full \wvn s
$\psi({\phi})$
in the ``position coordinate'' $\phi$, all eigenstates of the \sq
\h are
potentially present in the \wv and in the \dm  $\rho(\phi',\phi)$. 
Therefore, to compare with the results of the previous section we 
 must specify 
some basis of \wvn s and  evaluate the matrix elements of
$\rho(\phi',\phi)$ in that basis. We shall use either
the  two lowest energy eigenstates, or  the ``up'', ``down'' basis
corresponding to the eigenstates of $\sigma_z$.

\begin{figure}[h]
{{\includegraphics[width=0.5\hsize]
{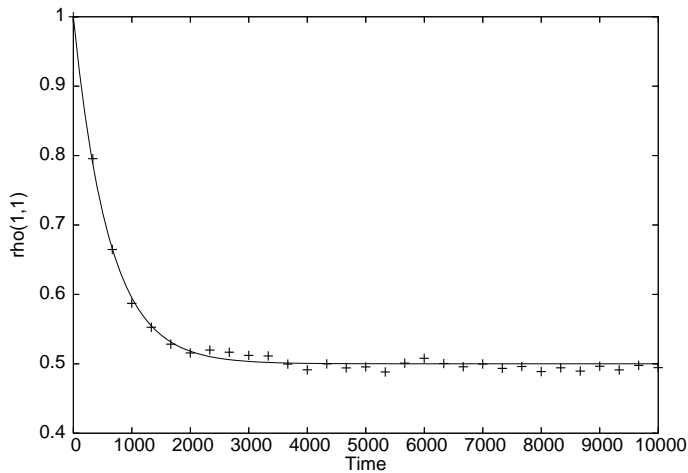}}}
\caption{Effect of  decoherence, as simulated by random  noise in
the full \h  Eq~\ref{hama3}, for the symmetric case with \px =0.
\sq parameters were $\beta=1.19, \xi=16.3$. The initial state is
the lowest energy eigenstate.
The pure state at t=0 is converted to the maximally mixed state. 
The quantity plotted is $\rho_{11}$ in the basis of energy
eigenstates, which  for \px=0 corresponds to $\frac{1}{2}(1+P_x)$.
The
 data is  fit to $\frac{1}{2}(1+e^{-Dt})$ yielding
$D=0.0017$, in good agreement with the prediction  D=0.0018 from
Eq~\ref{d21a}. Noise parameters
were $\Delta=0.00032, \omega_c=0.05$.  With the ``typical''
time unit $({L/pH ~C/pF})^{1/2}\times ~1.0\times 10^{-12}$
seconds$=6.3\times 10^{-12}$seconds,  the decoherence time
$1/D=560$ corresponds to 3.5 ns.  }
\end{figure}

\subsection{ Damping of  ${\bf P}_T$}
In general the evolution of the density matrix will reflect the
simultaneous effects of the internal \h  and the external
interactions.
 In 
Eq~\ref{pdota}, D gives a damping of the transverse components of
$\bf P$. Thus  the
simplest way to see the effects of the \de alone is to start from 
  an eigenstate of the symmetric,  \px=0, \h where $\bf V$ is
parallel to $\bf P$ and purely transverse.  In the ``spin 1/2
picture'' $\bf P$ would  start in the x direction with value 1 and
 decay exponentially  with decoherence time $1/D$  to the value
0.5.

For the numerical simulation of this situation with the full \sq \h
we begin with the lowest energy
eigenstate, obtain the \dm at a certain time, and take its
expectation value with respect to the starting \wvn, which quantity
we call $\rho_{11}$. 
Fig.\,4  shows  the results using the
noise parameters $\Delta=0.00032, \omega_c=0.05$ for $
\cal N$. An
exponential fall-off is observed, and a  fit gives $D=0.0017$. From
Eq~\ref{d21a} with  $\phi_c\approx 0.90$ we predict $D=0.0018$.
This is in good agreement with the fit. 
We note that this value of $D$ is not large in comparison  to the
energy splitting $V_x=0.0044$, so that in 
Eq~\ref{pdota} both terms on the right are of the same order of
magnitude. A prediction of the curve from Eq~\ref{pdota} using this
$D$ and $V_x$ is essentially indistinguishable from  the fit curve.
Excitations to higher states
were allowed by the program (``second cut'' =4), but none are
evident in that
the decay of the curve is to $0.5$ and not a smaller value.
Runs with  two or four states for the ``second cut''
 showed no significant differences.

\subsection{ Damped Oscillations}
To exhibit the characteristic two-state oscillations,  Fig.\,5
gives the results of a simulation run with the same parameters, but
now with the
initial state chosen to be an eigenstate of $\sigma_z$, i.e
localized in one of
the potential wells. The eigenstate of $\sigma_z$ is found by
constructing   the 2x2 matrix for $\phi$ in the two first energy
eigenstates and then finding the eigenvectors of this matrix.

We anticipate that the \wv will
 oscillate back
and forth with
a damping governed by $D$, as is indeed seen in  Fig.\,5. The
quantity 
plotted is the density
matrix element with  respect  to the starting state, which has the 
 two-state interpretation  $\rho_{11}=(1/2)(1+P_z(t))$. Again
we find no perceptible difference between runs using a two-state
and a four state basis for the time evolution, indicating little
excitation of higher states with these parameters. 

\begin{figure}[h]
\centerline{{\includegraphics[width=0.5\hsize]
{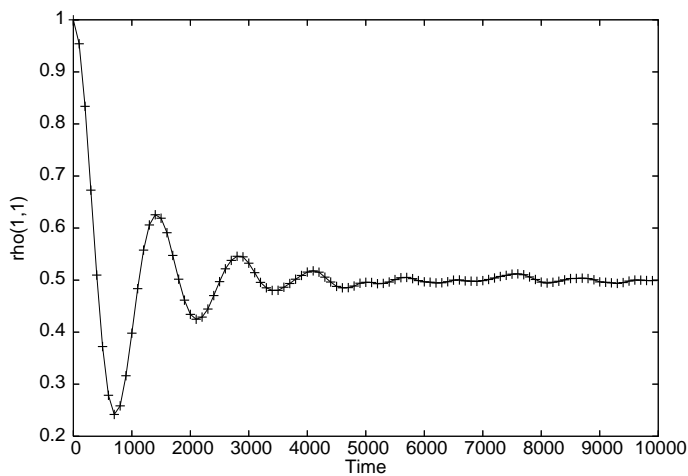}}}
\caption{Same conditions as Fig.\,4 but  with the initial state 
chosen as an eigenstate of $\sigma_z$. Since this is  not an
energy eigenstate oscillations occur, which  are then damped by the
decoherence. 
 The vertical axis corresponds to
$\frac{1}{2}(1+P_z)$. The curve is to guide the eye.}
\end{figure}

\subsection{ Turing-Zeno-Watched Pot Effect} 

 We briefly look at
  strong damping, which is the limit
\begin{equation}\label{str}
\frac{D}{\omega_{tunnel}}>>1\,.
\end{equation}
 Note that  in dividing 
Eq~\ref{pdota} by $V$, one obtains an equation containing only the
scaled time variable $tV$ and the parameter $D/V$, so that when
using a time variable  appropriately scaled to the oscillation
frequency, D/V is the only parameter in Eq~\ref{pdota}. With large
$D/V$ we     enter the regime of the ``Turing-Zeno-Watched Pot
Effect'' where
one  expects \cite{us} that the damping D inhibits the tunneling
strongly; if the \sy is in one of the potential wells it tends to
remain there. We approach the ``classical'' situation where quantum
mechanical linear combinations cease to exist.
   Fig.\,6 presents some simulations of
this situation. 
The conditions are as in Fig.\,5, but with  increasing
$\Delta$. Runs with
other oscillation frequencies, that is with different $\beta$,
yield
the same results when the time  is appropriately rescaled. 
 
\begin{figure}[h]
\centerline{{\includegraphics[width=0.5\hsize]
{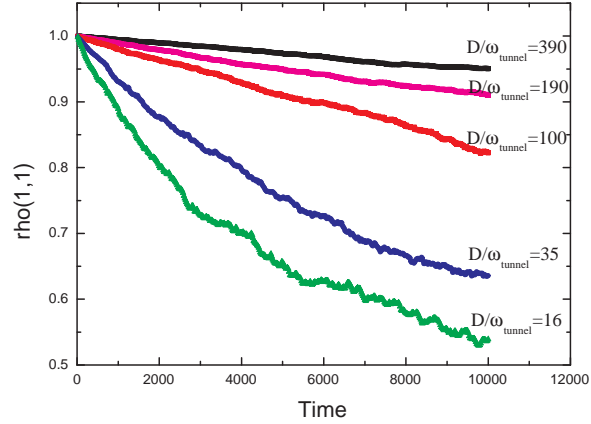}}}
\caption{  ``Watched Pot'' effect. Same conditions as in Fig.\,5
but with various degrees of strong decoherence
  ${D}/{\omega_{tunnel}}>>1$. The
 ``freezing'' of the evolution is
observed as predicted  \cite{us} from Eq~\ref{pdota} with 
 $\rho(1,1)\approx 1-\frac{1}{2}(\omega_{tunnel}^2/D)t$. The time
interval shown would correspond to about seven undamped
oscillations. The different values of $D$ were produced by varying
$\Delta$ from 0.01 to 0.002  and calculated from Eq~\ref{d21a}.}
\end{figure} 

\subsection{ Noise Frequency}
  According to
Eq~\ref{d21a},  changing 
$\Delta^2$ and $\omega_c$  such  that their
ratio remains
constant should leave D the same. However, 
 the higher frequency components of the noise spectrum  may  have
an independent effect
through the excitation of higher states. Excitations
beyond the lowest states
 would be undesirable, as it represents a non-unitary
evolution among the lowest states, with some probability going to
higher states.

 On the other hand, if the frequency of the noise is much less than
the
frequency corresponding to the distance to the next set of levels,
the adiabatic theorem tells us that the states remain in the lower
set and that such non-unitary effects are suppressed. Since our
principal
level splittings  are  generally  substantial fractions of unity,
 there should be little
excitation of higher states for noise frequencies
$\omega_c<<1$.
 In a plot of the
type Fig.\,4,
excitation of higher states is manifested by the decay
of the density matrix element to a value less than 1/2, showing
 population beyond  the first two states.

To examine these points, we show in Table III a  series of runs
at constant $\Delta^2/\omega_c$, but different $\omega_c$.  The
resulting D  as determined from a fit
as in Fig.\,4 and the final $\rho(1,1)$ are shown.
The approximate constancy of D seems to be well verified. For
$\omega_c<<1$
the decay is indeed to 1/2, but for larger values there are
noticeable departures. In  another set of runs with $\Delta$
reduced by a
factor of two this effect is almost entirely absent. With our
typical time
unit 
 these values of D corresponds to a \de time of $(1/D)\times
6.3\times 10^{-12} s=
3.5 ns$ and 14 ns. It should be noted that our noise spectrum
 has a
relatively
strong high
frequency tail $\sim 1/\omega^2$ as opposed to the exponential    
 cutoff one would expect for a purely  thermal background. 

The actual value of D is of course our great unknown, and we treat
it as simply a phenomenological parameter. For
orientation we can keep
in mind the estimate $D=T/(e^2 R)$ which we have used previously;
or  measurements by direct observation of the damped oscillations 
\cite {mooj} on a similar
\sq \syn. These find a \de time $\approx 20$ ns at 25mK. This
corresponds to the reasonable value $R=260\,k\Omega$  for the 
effective resistance. Our sample values of
$t_{dec}=1/D\sim
10^{3}$ in the dimensionless units, 
 with our typical time  unit of $6.3 \times 10^{-12} sec$,  would
correspond to some nanoseconds or tens of nanoseconds for
$t_{dec}$.

\section{Decoherence in the  NOT gate }

Having checked that our noise/\de simulation has reasonable
features, we now turn to some  applications.
The one variable or one-qubit problem is the simplest situation. 
When \px\,is swept (from a relatively
large value in the sense $V_z>>V_x$) to its opposite value, 
interchanging ``up'' and ``down'', it
represents the logic gate NOT. The understanding of the
effects of \de here is important both for 
finding the regime of operation of the logic gate and for  our
 proposal \cite{deco} for measuring D  via the success or failure
of
an \ad inversion.

\begin{table}
\begin{tabular}{|l|l|l|l|l|} 
\hline
$\omega_c$& $\Delta$&D, fit&D, Eq~\ref{d21a}&final $\rho(1,1)$\\
\hline
\hline
2.0&0.0020&0.0017 &0.0018&0.35\\
\hline
0.50&0.0010&0.0017 & &0.40\\
\hline
0.11&0.00045&0.0018& &0.48\\
\hline
0.050&0.00032&0.0018&&0.50\\
\hline
0.025&0.00023&0.0019&&0.50\\
\hline
\end{tabular}
\caption{ Values of D and final $\rho(1,1)$  for runs as in
Fig.\,4,  with varying  $\omega_c$ but constant
$\Delta^2/\omega_c$. The prediction of an
approximately constant D, as well as the value  for D, are  in 
agreement with  Eq~\ref{d21a}.
 The relaxation to a value below 1/2 for the
larger  $\Delta$ and $\omega_c $ values indicates excitation of the
third and fourth levels.}
\end{table}
In some applications of the \ad idea in quantum 
computing a special role is assigned to the ground state, as in the
search for the minimum of a complicated functional\cite{fahri}.
However  our
simple gates are not of this type and both states of the qubit,
either the ground state or the first excited state, are on an equal
footing. Thus in the NOT operation, for example, it is equally
important that  $1\to0$ or $0\to 1$, and which of these is
represented by the ground state is of no particular significance.

As in the  discussion concerning  \adyn, we 
measure the ``success'' of
a sweep by the value of ``final overlap'' as in Fig.\,3, where now
there is an average over realizations of the random noise.  Table
IV shows the effects of increasing sweep time with fixed noise
parameters. The sweeps are sufficiently long, according to Table
II, that non-\ady should be unimportant.

Interestingly,
although the noise parameters have been chosen to give a \de time 
of 29 000 time units, Table IV
 shows  \de  not fully setting in until significantly
longer sweep times. This  may be understood in terms of our 
remarks in connection with Eq~\ref{pdotb} 
 that  \de has most of its effect during level crossing.
If one attempts to estimate the time spent during the sweep in the
vicinity of the ``crossing'' with the given conditions, it is on
the order of 10\%. This
 suggests that
the relevant  time for the onset of significant \de is not 29
000 time units but rather more like 290 000, in agreement
with the behavior in Table IV. This argument is supported by
examination  of Fig.\,3 or Fig.\,13, where one sees that the actual
switching of a state takes place in a small fraction of the total
sweep time.
 Hence for our sorts of logic gates,
the speed needed to avoid \de may be less stringent than one might
simply infer from  comparing  the sweep time to the  \de time.

\begin{table}
\begin{tabular}{|l|l|l|l|l|l|l|l|} 
\hline
$\beta$&$\xi$& initial \px&$\omega_c$&$\Delta$
&$t_{sweep}$&final~overlap\\
\hline
\hline
1.19&16.3&0.0020&0.042&0.000042 &30 000&0.95\\
\hline
&& &&&60 000&0.96\\
\hline
&& &&&80 000&0.90\\
\hline
&&&& &150 000&0.85\\
\hline
&& &&&800 000&0.60\\
\hline
&& &&&1 000 000&0.58\\
\hline
&& &&&2 000 000&0.52\\
\hline
\end{tabular}
\caption{Effect of \de on the ``success'' of  \ad inversion sweeps.
Noise
parameters are chosen  to give a \de
time $t_{dec}=1/D= 29 000$  time units. ``Final overlap '' is the
average of the
overlap squared
 $\frac{1}{N}\Sigma_a \vert <\psi^a|\psi_f> \vert^2$ or
equivalently
$<\psi_f|\rho|\psi_f>$. As the sweep is made slower, so that \de
has time to
take effect at level crossing, $\rho$ approaches the totally
incoherent value of 1/2.  Errors on ``Final overlap '' are on the
order of a few percent. For the no-\de situation see Fig.\,3 and
Table II.}
\end{table}

\begin{figure}[h]
{{\includegraphics[width=0.7\hsize]
{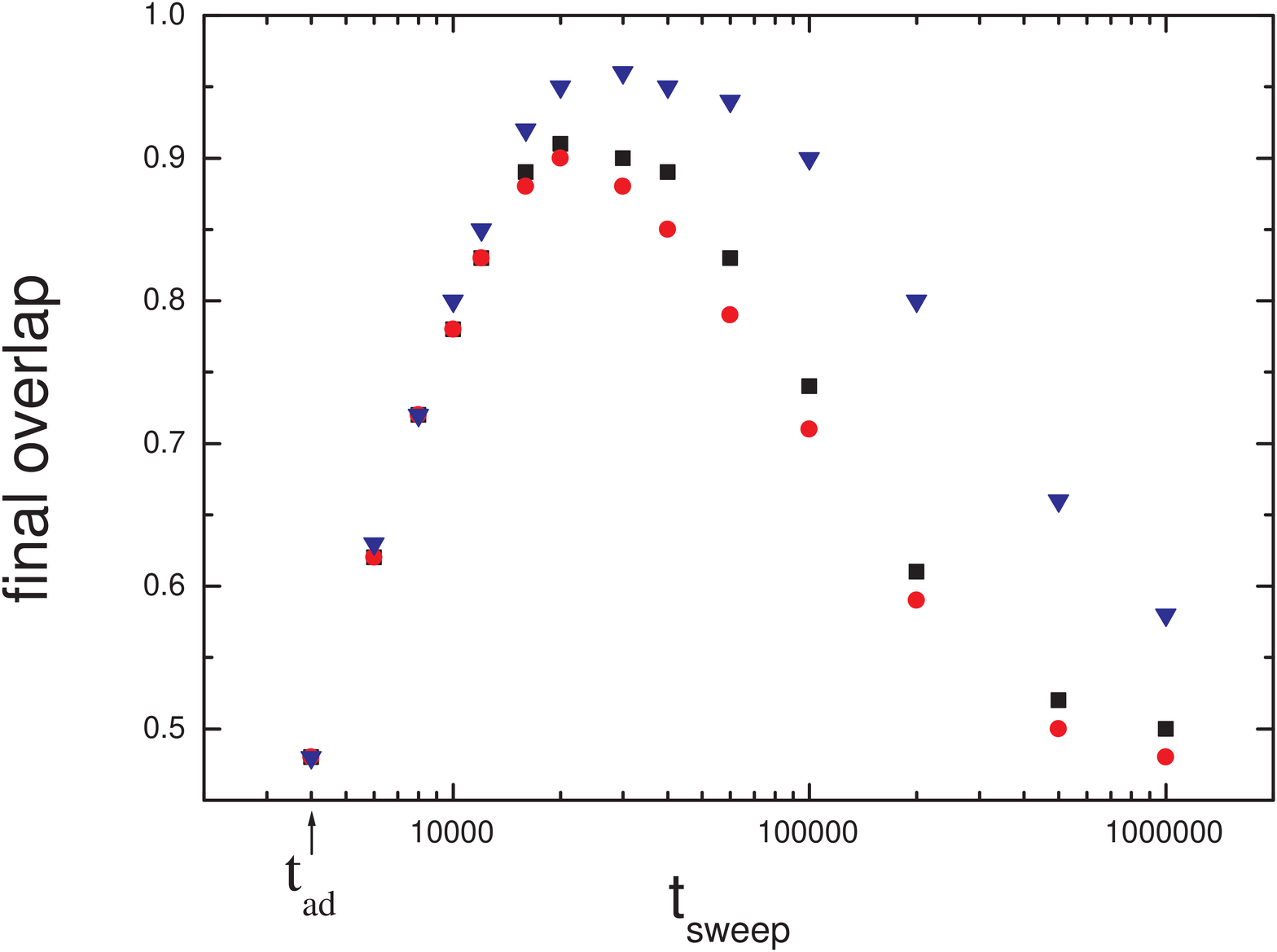}}}
\caption{Adiabatic and \de effects on the ``success'' of a sweep
($-0.002\to 0.002$) in \px\,, for the one dimensional \sy (NOT) as
the
sweep time is increased.
The \ad time for 50\% ``success'' is indicated by an arrow. As the
sweep time is lengthened one observes  the onset of \adyn, followed
by
decreasing ``success'' as the \de  takes effect. 
Squares and dots: noise parameters adjusted to give 
$D=1.3\times10^{-4}$. 
Squares: $ \omega_c=20mK$ and  $\Delta=5.0\times
10^{-5}$. Dots:
 $\omega_c=50mK ;~\Delta=7.9\times 10^{-5}$. 
Triangles: noise parameters adjusted to give $ D=3.5\times10^{-5}$,
using
$\omega_c=50mK$ and $\Delta=4.2\times10^{-5}$. 
\sq parameters are $\beta=1.19,\xi=16.3$.}
\end{figure}

\begin{figure}
{{\includegraphics[width=0.5\hsize]
{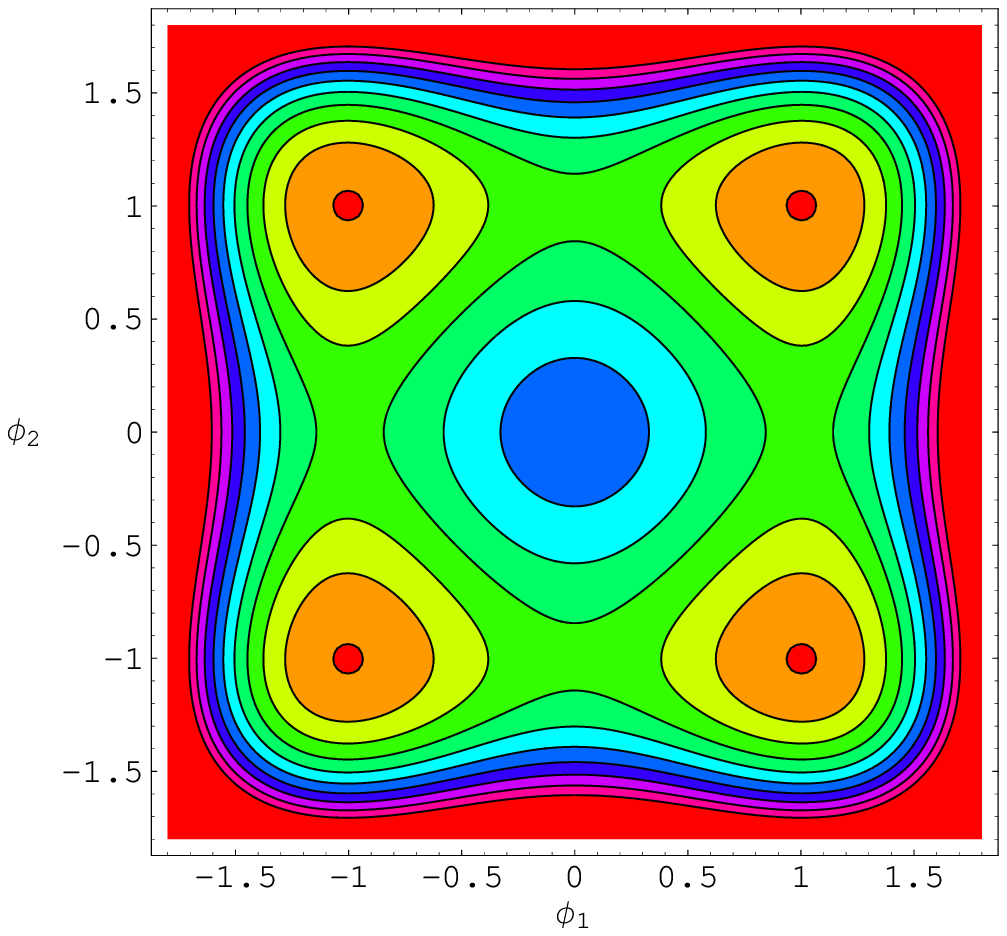}}}
\caption{ Potential contours for Eq \ref{v}, showing the  four
minima of the 2-qubit \syn.}
\end{figure}

To illustrate the effects of non-\ady and \de together, we show in
Fig.\,7 a series of simulations  for varying sweep times and
$\Delta$'s. There
appears to be a broad
range around twenty thousand time units where
neither effect is drastic.
The times refer to
our dimensionless units, so for our  typical  time unit
 $ 6.3\times 10^{-12}s$,   the two values of D in the plot would
 correspond  to
 $ 1/D=t_{dec}=$50 ns and  180 ns.
 If one were to consider \sq parameters giving a larger
$\omega_{tunnel}$ and so a shorter \ad time, the favorable region
could be widened considerably to the left, i.e. to shorter times.
(See the discussion below ``Smaller $\beta$'')

\section{Two variable \sy}

We now turn to the two variable or two-qubit \syn. When
two   one variable 
\syn s   are weakly coupled we arrive at a two
variable \sy with four low lying states. For the \sq these would
be the four possible states arising from the current circulating
clockwise or counterclockwise in two \sqn s, and  as explained in
ref~\cite{pla}
this
can be so arranged  that  the result of the sweep of one of the
\px depends on the state of the other \sqn, thus providing the
conditions for a CNOT gate~\cite{ave}. We recall \cite{ref1}
\cite{pla} the \h for this problem
\begin{equation}
\label{hama}
H={-1\over 2\mu_1}{\partial ^2\over\partial \phi^2_1 }+{-1\over2
\mu_2}{\partial ^2\over\partial \phi^2_2 }+V\;,
\end{equation}

with $V$ 
\begin{equation}\label{v}
V=V_0\{{1\over 2}[l_1(\phi_1-\phi^{ext}_1)^2
+l_2(\phi_2-\phi^{ext}_2)^2-
2l_{12}(\phi_2-\phi^{ext}_2)(\phi_1-
\phi^{ext}_1)]+\beta_1cos\phi_1+\beta_2cos\phi_2\}\; .
\end{equation}

In place of  $\mu=V_0$ for  the one variable case, one now has 
$\sqrt{\mu_1\mu_2}=V_0$. Analogously to the single variable case,
the factor $E_0$ which converts the energy of the dimensionless \h
to physical energy involves the inductance and capacitance of the
two \sqn s \cite{ref1}, namely $E_0=1/\surd LC$, where 
${1\over
L}={\sqrt{L_1L_2} \over
L_1L_2-L^2_{12}}$ and $C=\sqrt{C_1C_2}$. The small $l$ are the
inductances refered to $L$. In Fig.\,8 we reproduce a contour plot
for the potential showing its
four minima, corresponding to the four states of the 2-qubit \syn. 

 As in the one dimensional case, the numerical calculations use a
large harmonic oscillator basis, now in
two variables. For most runs a basis of one or two thousand
states was used. Even when using the ``second cut'' reduced basis
to four eigenstates,  time dependent runs with many samples could
take several minutes on  fast  PC's. It   seems that extensions to
more than two or three variables would need new computational
methods.

A basis
for the four states is provided by
``spin
up(1)spin down(2)=$u_+(1)u_-(2)$ and so forth:
\begin{equation}\label{base}
u_-(1)u_-(2)~~~~~~~~~~~~~~~~~~u_+(1)u_-(2)~~~~~~~~~~~~~~~~u_-
(1)u_+(2)~~~~~~~~~~~~~~~u_+(1)u_+(2)\;.
\end{equation}
These four states correspond to  definite senses for the currents
in
the \sqn s.

 On the other hand if the  parameter $l_{12}$
giving the flux coupling between the \sq is very small, the
situation
reduces to  two independent  single variable \syn s. Thus 
for the   analysis of weak coupling with small $l_{12}$  
it is convenient to introduce the eigenstates for any value of the
individual \px 
\,for each \sqn :

\begin{equation}\label{basev}
v_-(1)v_-(2)~~~~~~~~~~~~~~~~~~v_+(1)v_-(2)~~~~~~~~~~~~~~~~v_-
(1)v_+(2)~~~~~~~~~~~~~~~v_+(1)v_+(2)\;.
\end{equation} 
This is the independent \sq basis, where each \sq can have its own
\hn, according to its \px.
 
The $v$ are those  discussed in
 connection with Table I: $v_+=(cos \frac{\theta}{2}\, u_+ +
sin\frac{\theta}{2}\, u_-)$ and $v_-=(cos \frac{\theta}{2}\, u_- -
sin\frac{\theta}{2}\, u_+)$ . If the applied \px are different,
then we have different angles $\theta_1$ and $\theta_2$ in these
relations. 
At $l_{12}= 0$ the $v(1)v(2)$'s are the
eigenstates of the complete \syn,
 and with  $l_{12}\neq 0$ there will be a mixing among them.

\section{coupled \h in the spin picture}
   In the spin picture for Eq~\ref{v} there are
 now  two  ``spin 1/2'' objects, interacting through $l_{12}$
and subject to the  external fields $\phi^{ext}_1,\phi^{ext}_2$.
By the arguments used for Eq~\ref{ax}, we make the identifications

\begin{equation}\label{phid}
\phi_1\to\phi_c(1)\sigma_z(1) ~~~~~~~~~~~~~~~~~~~~~~             
\phi_2\to\phi_c(2)\sigma_z(2)
\;,
\end{equation}
and  the effective spin \h is
\begin{equation}\label{hab}
\tfrac{1}{2}\,{\bf \sigma}(1) \cdot {\bf V}_1+\tfrac{1}{2}\,{\bf
\sigma}(2) \cdot {\bf
V}_2 - V_0l_{12}\biggl(\phi_c(1)\sigma_z(1)-
\phi^{ext}_1\biggr)\biggl(\phi_c(2)\sigma_z(2)-\phi^{ext}_2\biggr
)\;.
\end{equation}

\begin{figure}
\begin{center}
\begin{tabular}{cc}
{{\includegraphics[width=0.5\hsize]
{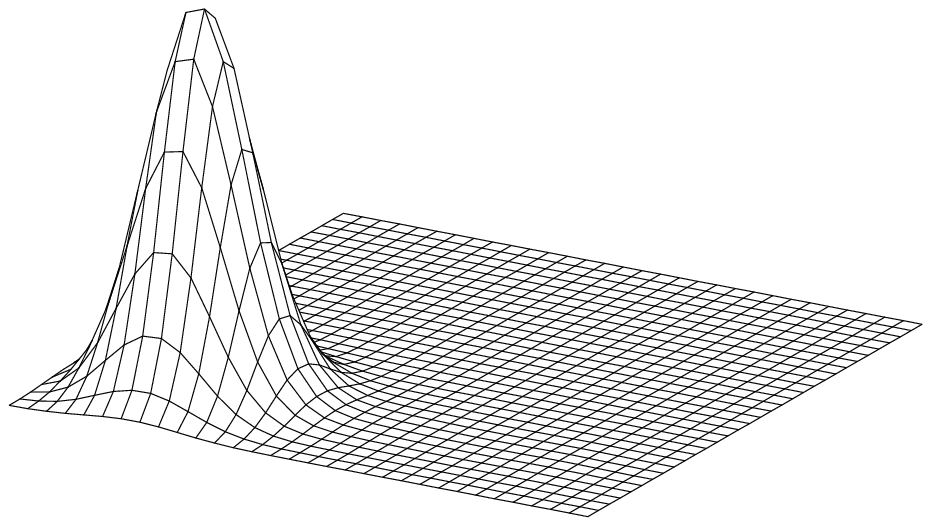}}}&{{\includegraphics[width=0.5\hsize]
{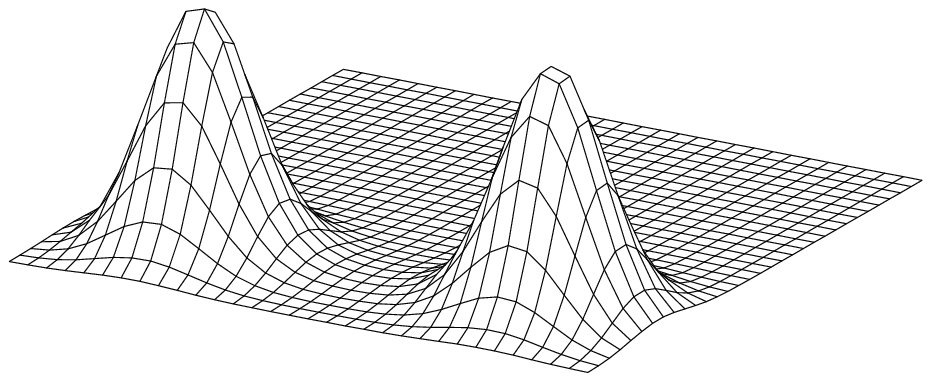}}}
\end{tabular}
\end{center}
\caption{Left: Square of the \wv for state 1 localized in the lower
left
potential well of Fig.\,8. Right:  at ``level crossing'' as the
\wv
moves to the lower right potential well. Potential
parameters  were $l_1=l_2=1,l_{12}=0.005$,
$\beta_1=\beta_2=1.19$  and $V_0=16.3$.}
\end{figure}

For small $l_{12}$ the components of the $\bf V$ are determined as
in Eq \ref{est}, namely $ V_z \approx 2 V_0\, l\, \phi_c 
\phi^{ext} $
while 
$V_x$ is found via Eq \ref{spl}. As before, $\phi_c$ is
approximately the  $\phi$ of a state
localized in one  of  the potential wells. 
The effective interaction \h
between the two devices is then 
\begin{equation}\label{hint}
H_{int}\approx -V_0l_{12}\phi_1\phi_2\approx
-V_0l_{12}\phi_c(1)\phi_c(2)\sigma_z(1)\sigma_z(2)\;.
\end{equation}
This operator  induces  level shifts of the four $u$
base states without mixing them. However when the tunneling is
introduced as a perturbation, non-trivial combinations of the
four states can arise.

\subsection{Low-lying level patterns}

We begin with weakly interacting \sqn s, $l_{12}<<l_1,l_2$.
The energy level pattern anticipated for the lowest levels  may be
understood by beginning with the totally
decoupled \sy of just two independent devices, $l_{12}= 0$.
If we   take both devices
with about the same parameters, for example, we have 
 first the ground state where both \sqn s are in their lowest
state, the first state of Eq~\ref{basev}. Then there are two
approximately  degenerate states with
one \sq in the first excited state and the other in its ground
state, the second and third states of Eq~\ref{basev}. Finally the
fourth state has both \sqn s in the first
excited states, the last \wv of Eq~\ref{basev}. If each device is
in the configuration where the
splitting of its first two states is small (small $V_z$),  as in
our
discussions above, then the splitting from the ground state to the
degenerate pair is equal to the splitting from the pair to the
fourth state.  Furthermore the  splitting to the fifth state 
should be substantially greater since it involves an excitation of
the principal quantum number. 

Turning on  a small $l_{12}$, we show 
an example from numerical calculations in Table V.   With
$l_{12}=10^{-5}$ these
reveal the expected pattern.
  One sees that the fifth state is 
 well separated from
the lower ones,  again supporting the use of the picture of an
approximately isolated Hilbert space, as for the first two
states
in the single variable case. 

\begin{figure}
\centerline{{\includegraphics[width=0.7\hsize]
{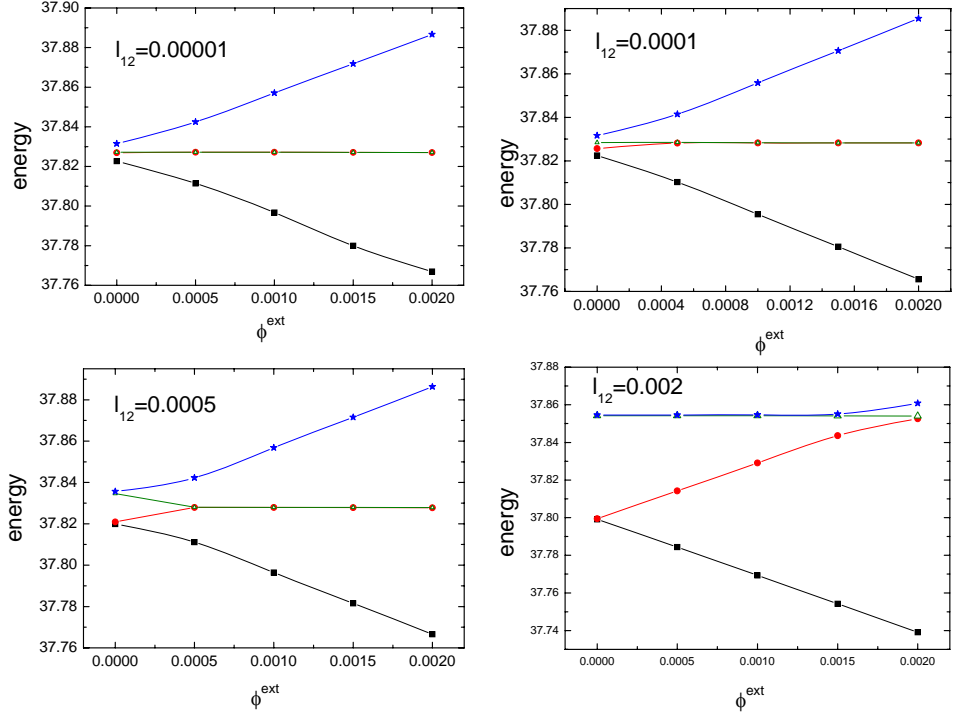}}}
\caption{Level behavior for  the first four states of two identical
SQUIDs  with coupling $l_{12}$. In the first panel one notes the
pattern of a
ground state, an almost degenerate pair and single state, as would
be expected for two independent \sqn s.  This pattern changes as
the interaction energy between the SQUIDs $\sim V_0 l_{12}$ becomes
on the order of
the splitting $\omega_{tunnel}$. }
\end{figure}

In the example of identical parameters for both \sqn s, with
second
and third levels degenerate for $l_{12}=0$, we may use
degenerate 
perturbation theory to find the splitting induced by turning on
$l_{12}$. The matrix element is
$<v_+(1)v_-(2)|H_{int}|v_-(1)v_+(2)>=
V_0 l_{12}(sin\theta)^2$, where $\theta$ is the angle for going
from
the
$u$ to the $v$  as discussed  in connection with Table I.
The splitting of levels 2 and 3 is then 
$2 V_0
l_{12}(sin\theta)^2\phi_c^2$. With our typical parameters and $
l_{12}=1\times
10^{-5}$ and  
 \px =0 and so $sin\theta=1$, the formula, using $\phi_c\approx
0.88$,  gives $25 \times 10^{-5}$. The numerical calculation, as
shown in
Table V, gives $27 \times 10^{-5}$.

 Observation of this small splitting would be quite amusing with
regard to the question of coherence between macroscopic objects.
While work with the ``Cooper-pair box''\cite{pash} has seen effects
involving
interference between the states of two qubits, the states involved
differ only microscopically, namely by a Cooper-pair. Here, with
the \sqn,
the states  concerned differ by the circulation direction of a
macroscopic number of electrons, and so seem to pose the
macroscopic coherence question more dramatically. 
  The energy eigenstates resulting from the diagonalization to
obtain the small splitting are 
$(1/\surd2)\bigl(v_-(1)v_+(2)\pm v_+(1)v_-(2)\bigr)$. The
splitting, one might say, results from the relative quantum phase
$\pm$
involving the two \sqn s, that is, between two macroscopic objects.
This is yet a step further than the effects for one \sqn , where
one is sensitive to the phase between different macroscopic current
directions, but  in one device.
Since this splitting is very  small, however,  line broadening  due
to the
 noise effects may be comparable to the splitting. Values of
$D\sim 10^{-4}-10^{-5}$ for the tens of mK region could be in the
same range as the splitting for $l_{12}\sim 10^{-5}$.

The degenerate perturbation theory  formula only applies when we
have the
picture of two approximately degenerate levels well separated from
the others, as in Table V. The picture changes rapidly as the
mutual coupling $l_{12}$ is increased.
In Fig.\,10 we show the energy level patterns for the first four
levels for different  $l_{12}$ as functions of a common applied
\px. Again, the two \sqn s are taken to be identical, with our
standard parameters. One observes   that after about  $l_{12}
\approx  10^{-4}$ the uncoupled \sq pattern no longer holds, since 
now the interaction  energy $\sim V_0 l_{12}$ is on the same order
as the original splittings, $\omega_{tunnel}=0.0044$. A
characteristic change of behavior occurs when 
$l_{12}\approx$\px. This may be understood as the value of $l_{12}$
 where the
flux contributed from the other \sq becomes comparable to the
applied \px.(see
Eq 5 of 
ref~\cite{pla}).

Another limit which is not difficult to analyze is that of small
$\theta$ or relatively large \px (see Table I). In this limit the
$u$'s are the eigenstates and the interaction Eq~\ref{hint} simply
gives additive contributions to the energies. We may even consider
different  parameters and \px\, for the \sqn s. According to
Eq~\ref{hab}  this results
in a ground state with both spins ``down'' and energy $\frac{1}{2}(
-V_z(1)-V_z(2)-\delta)$ and an upper state with both spins ``up''
and energy $\frac{1}{2}(
+V_z(1)+V_z(2)-\delta)$. There  are two middle states with energy 
$\frac{1}{2}(+V_z(1)-V_z(2)+\delta)$ and $\frac{1}{2}(-
V_z(1)+V_z(2)+\delta)$;  $\delta$ is the contribution from
Eq~\ref{hint}, $\delta=V_0l_{12}\phi_c(1)\phi_c(2)$. In the case of
identical \sqn s these middle states form a degenerate pair, as one
sees for the larger \px. With small deviations of the
$\bf V$ from the z-direction there will again be a splitting which
can  be found from perturbation theory.

\section{CNOT configurations}

Our design for a  CNOT operation   consists of an \ad sweep from a
point  with relatively large \px\, in the
$(\phi
^{ext}_1,\phi ^{ext}_2)$  plane to another such point
so that there is a definite mapping among the four $u(1)u(2)$
states.  The four possible
configurations of currents clockwise and counterclockwise in the
two \sqn s  undergo  a certain definite, reversible, rearrangement.
This rearrangement is chosen in accordance
with the definition of CNOT: one
pair of states remains unchanged (control bit is zero), while the
other pair reverses (control bit is one). This is accomplished  by
having the energy eigenstates (1,2,3,4), each one
concentrated in a different  potential well, move
from one set of locations   to another~\cite{pla}. 
Since for relatively large \px each well represents a distinct
state of the \sq currents, one physical configuration of  the two
qubits is mapped to another.
These rearrangements are
represented by ``tableaux'' like $\pmatrix{3&4\cr1&2}$ showing the
localization of the first four energy eigenstates   in the
potential
wells of Fig.\,8.

\begin{table}
\begin{tabular}{|l|l|l|l|l|} 
\hline
$\beta$&$\mu$
&$l_{12}$&\px&$\Delta~E$\\
\hline
1.19&16.3&0.00001&0.0&0.0043\\
\hline
&&&&0.00027\\
\hline
&&&&0.0043\\
\hline
&&&&0.42\\
\hline
\end{tabular}
\caption{ Splittings at \px=0 among the first five levels
for the first panel of Fig.\,10.  The  energy difference $\Delta E$
is with respect to the
previous level, thus  the first excited state is
0.0043 units above the ground state.  The fifth state is
distinctly separated from the first four. The first and third
 splitting should be
approximately that  of a single \sqn, which with \px=0 is 0.0044.
The small splitting between the second and third state is in
agreement with the prediction of degenarate perturbation theory,
$2V_0l_{12}\phi_c^2$. }
\end{table}

In a first step it is only necessary to identify those locations
of the $(\phi^{ext}_1,\phi ^{ext}_2)$ plane with different tableaux
such that a sweep from one to the other leads to the desired
rearrangement. An example is $\pmatrix{3&4\cr1&2} \to
\pmatrix{4&3\cr1&2}$.  The  lower row may be identified with
control bit  zero, and the upper row with control bit one. 
The adiabatic theorem then guarantees that a ``sufficiently slow''
sweep will preserve the occupation of the levels, $1\to 1, 2\to 2$
and so forth. Once having so
identified the desired sweep, the  \ady may be examined afterwards.

  Not all points of the $(\phi^{ext}_1,\phi ^{ext}_2)$ plane
are suitable starting or ending points for a sweep since we require
``well
defined'' wavefunctions.
The \wvn s should be A) well localized in  one potential well so
that the \sq is in a definite flux state, and B)
all first four states should be localized in different wells. 
Fig.\,9  shows the appearance 
of a  ``good \wv ''  with state 1 in the lower left potential well
of Fig.\,8. In the searches 
a  \wv was considered  as ``well localized'' when the distance
between
centers of the \wvn s was more than 2.5 times the spread of the
\wvn, as measured by the dispersion.

 In Fig.\,11 we show the results of such a search.
 The associated tableaux are indicated for each region. The black
areas are those of ``bad \wvs''. The general range of ``good''
corresponds to the observations for one \sq in Table I, where the
spin picture was valid up to about \px $\approx 0.01.$
   
In principle a CNOT may be accomplished by a sweep between any two
adjacent regions where one row (or column) interchanges and the
other does not.
As was explained in ref.\cite{pla},  the
switching  values between tableaux (dark lines) may be
understood  as a ``level crossing'' occurring when the flux from
the
control \sq onto the
target \sq is equal and opposite to the applied flux onto the
target \sqn. That is, when 
\px $=l_{12}/l$. At this point the total flux on the target \sq is
approximately zero and so  is at a level crossing. This argument
allows one to  understand the pattern of dark lines.
It will be noted that in crossing a dark line only adjacent energy
levels switch, as expected for a ``crossing''. The vertices
are singular points involving more than two levels.

 For a smaller $l_{12}$ the map will then be similar but with
smaller regions between the vertical and horizontal
 dark lines. In the central box of the plot, although there are
``good" \wvsn,  the applied flux is apparently too low for the
``immobilization'' argument of 
ref.\,\cite{pla} to work and there are  no CNOT rearrangements
within the box.
However CNOT sweeps  exist connected to peripheral regions. These
take place in the vertical or horizontal
direction with the flux on one qubit (control bit) relatively large
and constant while the other flux (target bit) is varied.

For such vertical or horizontal sweeps, involving changing one
flux,
  only one \sq inverts, according to which one undergoes the sweep.
 In diagonal sweeps involving both fluxes on the other hand, 
states transfer across the
diagonal in the tableaux, indicating changes in both \sqn s. This
explains
why the diagonal black lines are very narrow, since such ``double
flips''
involve two tunnelings with a corresponding small mixing energy. 
\begin{figure}[h]
\centerline{{\includegraphics[width=0.8\hsize]
{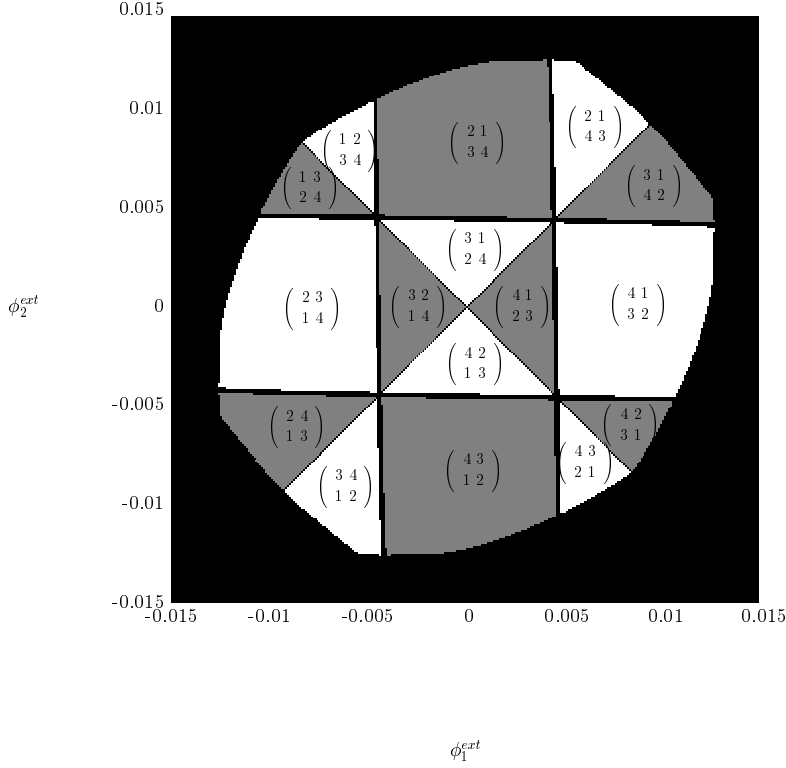}}}
\caption{Map of the regions of \pl plane with well defined
\wvn s suitable
for performing CNOT. The numbers in  a tableaux show  the location
of the 
first four energy eigenstates  on the potential
landscape
of Fig.\,8. These show that a CNOT may be obtained by horizontal or
vertical sweeps in the regions external to the central square.
Potential parameters used were $l_1=l_2=1,l_{12}=0.005$,
$\beta_1=\beta_2=1.19$  and $V_0=16.3$. }
\end{figure} 

\begin{figure}[h]
{{\includegraphics[width=0.8\hsize]{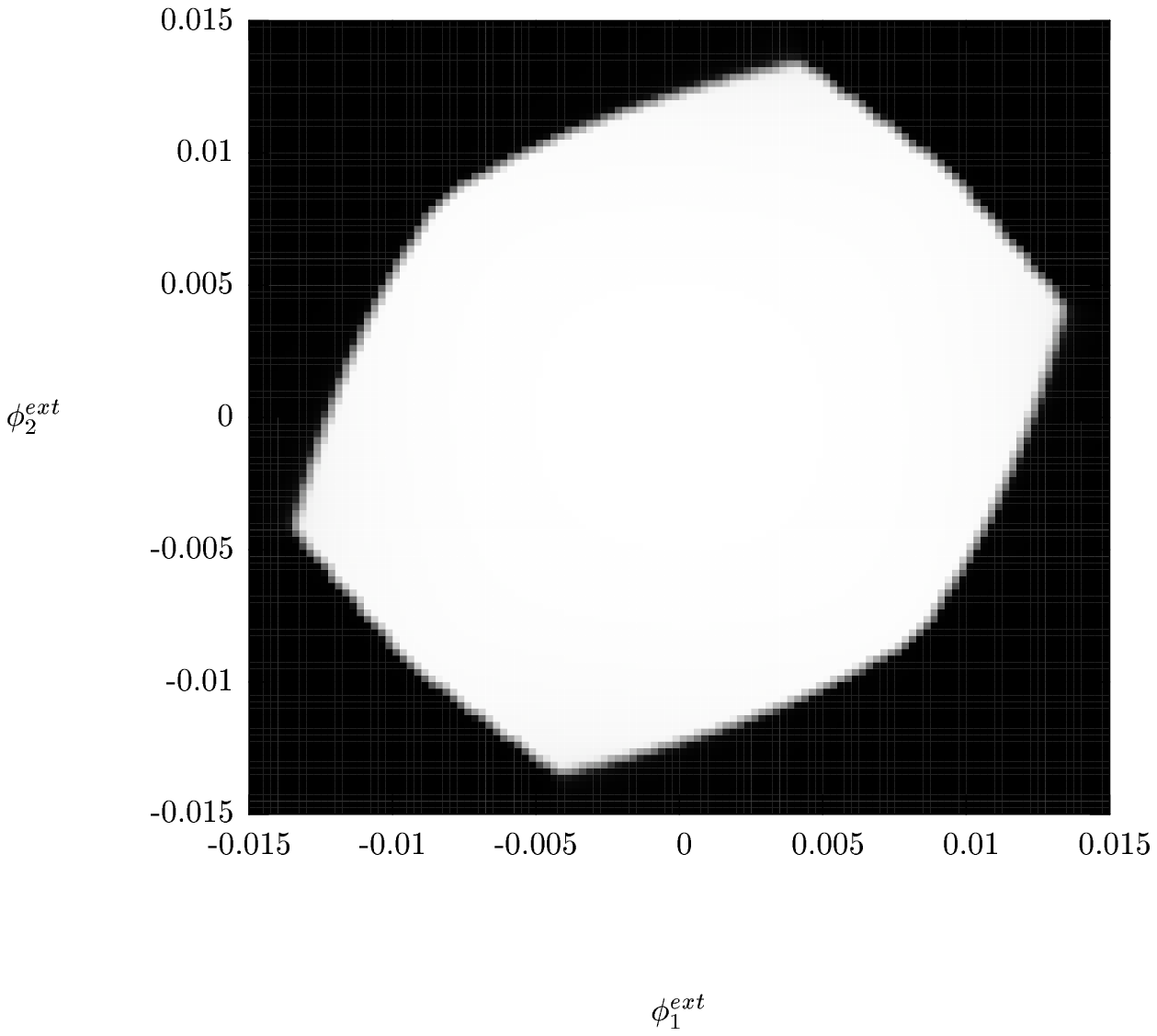}}}
\caption{Region of the \pl ~plane with  high ``completeness ''
using Eq \ref{mm}. One notes the close
similarity to the  region defined by  Fig.\,11, which was  found by
examination of  individual \wvsn . }
\end{figure}

Although the map of Fig.\,11  was obtained by an examination of the
individual \wvn s, a reasonable idea may be had
 by simply inspecting the potential landscape. Since the  ordering
of the energy levels will usually follow the ordering of the minima
of the potential wells, we  find that the ordering of the
minima usually give the correct tableaux. Thus
a suitable sweep is frequently simply one where the potential
minima rearrange  in the desired manner. Alternatively, all the
transition lines on the map may be found from the  set of 
linear relations arising from the ``level crossing'' conditions
(coefficients of the $\sigma_z$) while treating  the tunneling
($\sigma_x$) terms as a perturbation. 

\subsection{Hilbert space completeness}
For the ``spin 1/2 picture'' to be a meaningful in a 
complex \sy with many levels it is necessary that the states
selected constitute an approximately
 independent vector space.  If, for example, the selected states
were to mix with other states of the \sy as we carry out our gate
operations, then the two states  would not be a faithful
representation of a two-state qubit, since more than two states
would be involved. And similarly
when we have two qubits or  four states for CNOT, these four should
act effectively as a  separate space.

 The Hilbert space completeness test mentioned at the end of the 
section ``Identification with a spin 1/2 system'' proves to be
quite
interesting in this regard and may be a useful tool
in studying higher dimensional \syn s. In this method, we take a
point in the parameter space of the \h and compare it with some
reference point. If the Hilbert spaces for these different \hn s
are closely the same, there should be a high overlap of the
lowest eigenstates for the different \hn s. As we shall explain, a
test can formulated which applies to any linear combination. For
the one variable \sy
the parameter space in question
 is the  \px~line, and  with two variables 
  the \pl~plane. The precise  point we choose as the reference is
unimportant in regions where the overlaps are high; for the
\pl~plane we take \pl =(0,0).

Let the $v_i$ be the  lowest eigenstates at the reference point
  and  $v'$  a linear combination of the lowest eigenstates for
some other parameter values. We search for the ``worst case'', the 
minimum value of
$\Sigma |<v_i|v'>|^2$. 
  The procedure involves constructing the matrix
$M_{ij}=<v_{i}\vert v'_{j}>$. For the one \sq or two state \sy i
and j run from one to two;  in the two variable system   from one
to four.

One now observes that the sum of the overlaps squared
$\Sigma |<v_i|v'>|^2$  is the expectation of the matrix
$M^TM$ in the state $|v'>$. Thus the
worst or lowest overlap for an arbitrary normalized linear 
combination in the $v'$ space is given by the lowest eigenvalue of
this matrix:

\begin{equation}\label{mm}
least~overlap=smallest~eigenvalue ~M^TM  
\end{equation}

 Varying the chosen point in the \pl~ plane,
one maps out the  region of a common Hilbert space where the worst
overlap is close to one.
 Fig.\,12  shows such a map, produced with high resolution.  A grey
scale was used for the points, from white
for  $least~overlap = 1$  to black for $least~overlap
= 0$.  

It is quite interesting that the region of the \pl plane found by
this method is essentially the same as that found by the 
detailed examination of individual \wvs as in Fig.\,11.
This may perhaps be understood by noting that if we have  ``bad
\wvs '' with two states in one potential well,  this will have
a poor overlap with the ``good'' situation where each \wv is in a
different well. On the other hand   ``bad
\wvs ''in the sense that they are delocalized, as at a
``crossing'', will still give a high ``completeness''. 
The sharp transition, with
little grey area, seems remarkable. Although the corresponding
transition in the one dimensional case was rather abrupt, it seems
more so here, presumably the effect is multiplicative.
 Both Figs.\,11 and 12 contain 90 000 pixels,
 necessitating runs of many hours.

Fig.\,12 does not contain the narrow black bands of Fig.\,11, where
there are ``bad \wvs'' at ``level crossings''. This is because
these ``bad \wvs'', although delocalized in $\phi$ and so not
corresponding to a
definite state of the \sqn s, are still in the same  Hilbert space
as  the localized states. An example of such a state is shown in
the Right panel of Fig.\,9. It should also be noted that the
narrowness of the black bands in Fig.\,11 is in accord with our
earlier remarks concerning the small fraction of the time spent
near ``crossing''
during a sweep.

\begin{figure}
\centerline{\includegraphics[angle=-90,width=0.5\hsize]{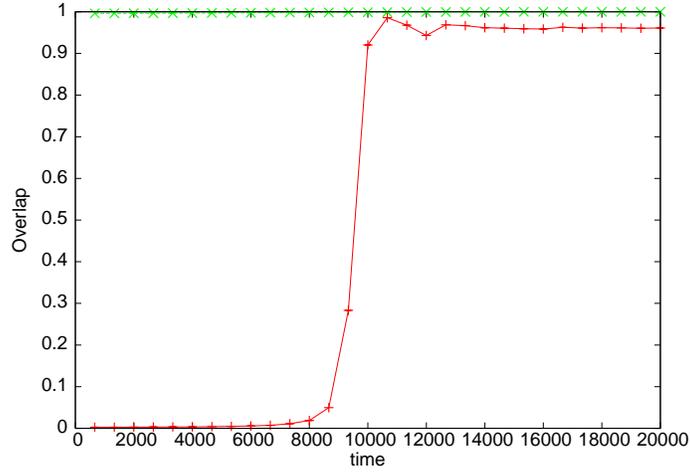}}
\caption{Behavior of the states 1 and 4 in a CNOT sweep connecting
two tableaux, where state 1 should  remain in its original
location while state 4 should reverse, as in the lower left of
Fig.\,11. No noise is applied. The vertical axis
shows the overlap of the \wv with its intended final state. State
1 (upper curve) should not change its location and  is seen to be
stable. State 4 (lower curve) should reverse its location and is
seen to go from  zero to approximately complete overlap with the
desired final state. State 2 and
state 3 are not shown but behave similarly to states 1 and 4
respectively. Squid parameters are  as in Fig.\,11, and the sweep
time was 20 000 units. It is to be observed that the switch itself
takes place in a small fraction of the sweep time.  }
\end{figure}

\subsection{ Noise for the two variable system}
 The extension of the previous  noise treatment to
the two variable problem is straightforward if we continue to
 represent the noise  as fluctuations of the external fluxes \px.
To each \px \, there will now be a noise term as in Eq~\ref{hama3}.
As in the discussion of Eq~\ref{hab} we have $ V_z \approx 2 V_0 l
\phi^{ext} \phi_c $
and so an additional term in Eq~\ref{hab} 
\begin{equation}\label{noises}
 -V_0\{l_1\phi_c(1) \sigma_z(1){\cal N}_1 + l_2 \phi_c(2)
\sigma_z(2){\cal N}_2 \}\; ,
\end{equation}
where again we keep only the linear term in the noise. 

There will then be a $D$ as in Eq~\ref{d21a} 
 associated with each variable
  
 \begin{equation}\label{d2}
 D_1=4 {(V_0 l_1\phi_c(1)\Delta_1)^2
\over \omega_c(1)}~~~~~~~~~~~~~~~~~D_2=4 {(V_0
l_2\phi_c(2)\Delta_2)^2 \over \omega_c(2)} \; .
\end{equation}

We can now carry out the simulations with the two noise signals
imposed, one for each \sqn.

\section{CNOT sweeps}
As for the one-bit case, the ``success'' of a sweep can be
measured by the 
overlap of an evolved \wv with the
stationary eigenstate for the intended final state. For CNOT two
\wvsn, representing control bit =0,   should not change their 
states,
while those representing control bit =1  should move to new
positions.
We exemplify a successful CNOT sweep in Fig.\,13 where we show the
behavior of states 1 and 4 for a sweep between a tableaux
$\pmatrix{3&4\cr1&2}$ and a tableaux $\pmatrix{4&3\cr1&2}$. One
sees the reversal of state 4 while state 1 remains steady.

\begin{figure}
\centerline{\includegraphics[width=0.7\hsize]{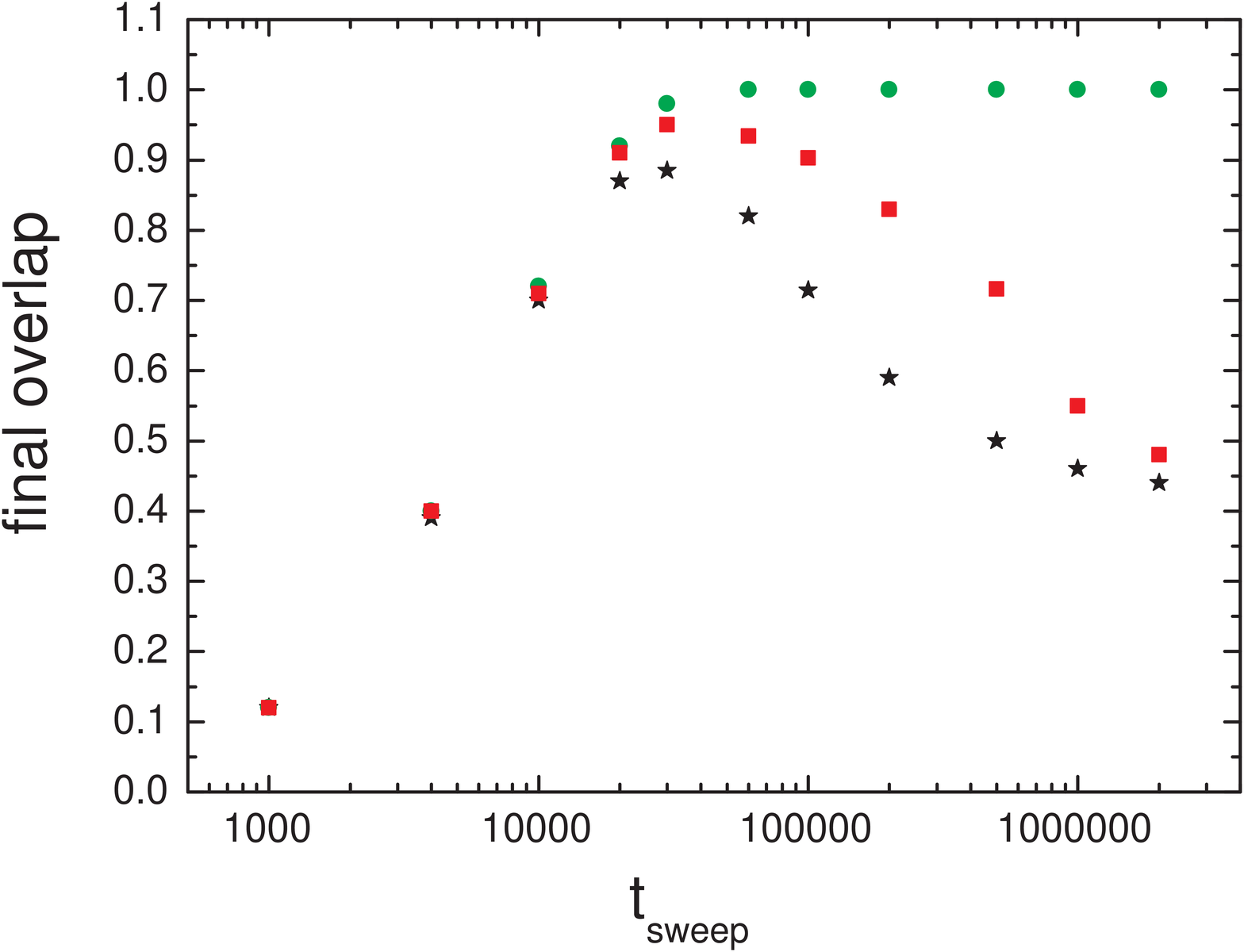}}
\caption{A CNOT sweep  in the lower right  of the \pl~ plane
of Fig.\,11, with  $(0,-0.01)\to (0.006,-0.01)$. The final overlap
is plotted  for state 1, which should change position. The coupling
between
the \sqn s  is $l_{12}=0.005$. The sweep without
noise (circles) shows the effect of non-\ady for fast sweeps, while
the addition of noise exhibit the noise/decoherence effects at
longer times.  The
squares have noise parameters $\Delta=0.000042,\, \omega_c=0.042$
and the stars  $\Delta=0.000079,\, \omega_c=0.042$. The noise is
applied to both \sqn s equally but independently.}
\end{figure}

In Fig.\,14 we illustrate the effects of non-\ady and \de for the 
$\pmatrix{4&3\cr1&2}\to \pmatrix{4&3\cr2&1}$
sweep of the right lower portion of  Fig.\,11, with  $(0,-
0.01)\to (0.006,-0.01)$. The curves show the behavior of the
overlap for state 1, which ideally would behave
as the switching curve of Fig.\,13, reaching close to complete
overlap. The upper curve, without
noise, shows the progression to \ady as the sweeps become slower.
The 
lower curves have  noise applied, equally to both \sqn s, to show
the  effects
of \den. The two sets of noise parameters, according to
Eq~\ref{d2},  
correspond to  \de times of 29 000 and 8 200 time units,  giving
\de times 1/D of  180ns and 52ns with our typical  time
unit.  In
terms of the formula $1/D=e^2R/T$, at 20mK  these would correspond
to  $R\approx 2 M\Omega$ and $R\approx 580 k\Omega$. 
As expected, the \wvs for the states which should not change
remained stable in these runs.

In these runs we applied the noise to each \sq
independently. However, one might consider the effects of a common
noise
applied to both \sqn s together.  
This would be the case not for the true intrinsic \de but could
represent, say, an 
instrumental effect due to an external  common noise  like long
range fluctuations 
in the applied field.  Interestingly, applying the noise signal so
it is
the same on both devices seems to have little or no effect.
Taking the lower curve of Fig.\,14 at $t_{sweep}=100\; 000,$ one
finds
that such a common noise produces the same result as the
independent noise,
namely a final overlap of 0.7. This at first surprising result is
understandable from the fact that the major effect of the noise
takes
place at level crossing. The noise is mostly effective on the
\sq undergoing the sweep and apparently the noise on the control
\sq has little effect. Indeed,  turning  the noise on 
\sq 2 completely off still leaves the final overlap at 0.7; but
turning it off for
\sq 1 leads to a perfect result with overlap $\approx$ 1.

\section{Smaller $\beta$}

As  seen in Table II, a reduced $\beta$
gives an increased $\omega_{tunnel}$ and shorter \ad times.
Since shorter times gives the \de less time to act, there
is reduced \den. This suggest examining the
situation with 
 a  reduced tunnel barrier. We take   $\beta=1.14$, which according
to  Table II, increases
$\omega_{tunnel}$ by a factor six, from 0.0044 to 0.027 energy 
units.
 Fig.\,15  shows, analogously to  
 Fig.\,11, the good domains of the \pl plane  for
$\beta=1.14$. The small change in $\beta$ has great effects. The
thin
black lines of Fig.\,11, with
 ``bad \wvs'', have now become broad bands.  Thus
starting and finishing points of the sweeps are now restricted to
certain ``islands''. By lowering the tunnel barrier, we have
created substantial areas   represented by the Right panel of
Fig.\,9,
  where the \wvs are not in a single
well with a definite flux in the \sqn s. While for
 Fig.\,11, this occurred only at the ``level crossings'' given by
the  black lines, there are now rather large regions with
delocalized \wvsn. 
 Even on the ``good'' islands  examination of pictures of the
\wvs as in Fig.\,9 shows that, while they are generally still 
well concentrated, there are little ripples some distance off from
the center of localization. Evidently in this parameter range we
operate on the  border to total delocalization.
  
\begin{figure}
\centerline{\includegraphics[width=0.8\hsize]{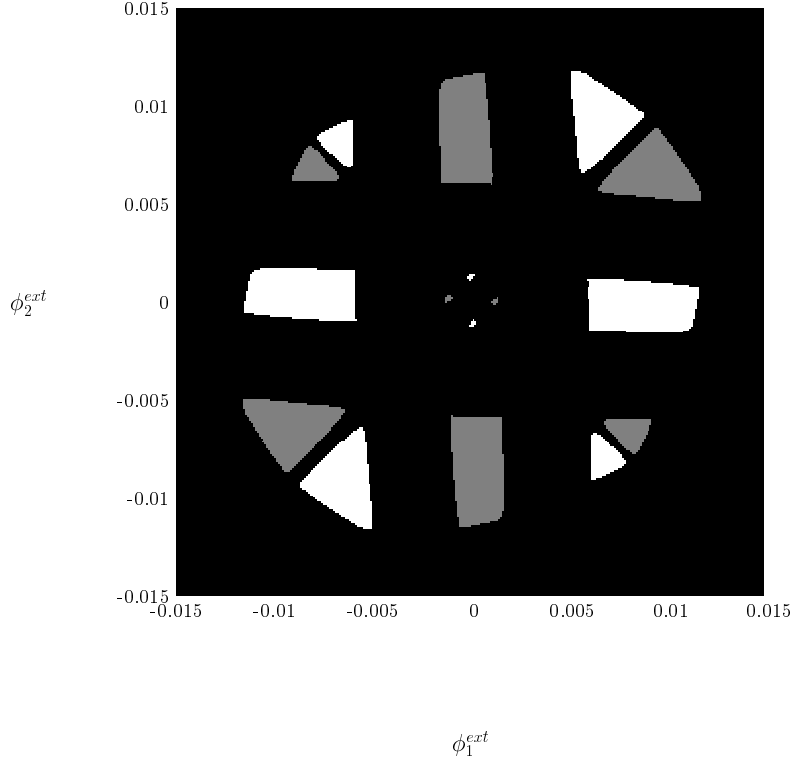}}
\caption{``Good domains'' of the \pl\, plane as in Fig.\,11  with
the smaller $\beta$ values $\beta_1=\beta_2=1.14$. Due to the
increased
tunneling the ``good''
regions where there is a definite state of the \sqn s have been
reduced  to small islands. The relative narrowness for the diagonal
bands,
as discussed for Fig.\,11, is seen here clearly. For legibility the
tableaux are not
shown; however they correspond to those on Fig.\,11. }
\end{figure}

To examine the hoped-for improved robustness\cite{robust} with
respect to
noise/\de we
show in Table VI  the results of a series of CNOT sweeps for
$\beta=1.14$, choosing  a sweep  on the lower left
of Fig.\,15 where $\pmatrix{3&4\cr1&2}\to
\pmatrix{4&3\cr1&2}$. The increased $\omega_{tunnel}$ allows us to
set
 $t_{sweep}$ as short as 1 000 units without violating \adyn.
Increasing the
noise parameter $\Delta$ from zero and tabulating  the final
overlap
for state
4, one sees that significantly shorter \de times as compared to 
$\beta=1.19$  (Fig.\,14) become possible.
   The next-to-last
entry would correspond to a \de time of only 500 time units. This 
reflects in part the phenomenon discussed in connection with
Eq~\ref{pdotb} and Table IV that the sweep time can be
 longer than the \de time before
large \de effects occur.

The 500 time units would be  3.1 ns with our
``typical'' parameters.
 This is  substantially less than the 20 ns
reported in  ref.~\cite{mooj}. This  is  one of our most
interesting
results concerning the engineering of such devices and perhaps
implies that the feasibility of such devices is not so very far
off. 

\begin{center}
\begin{table}
\begin{tabular}{|l|c|l|l|l|l|l|l|l|} 
\hline
$\beta$&$~~~~~~\mu,~~~~ V_0$~~~~ &$l_{12}$& initial \pl &
final \pl& $t_{sweep}$&state&$\Delta$&final~overlap\\
\hline
\hline
1.14&16.3&0.005 &(-0.0080,~~-0.0075)&(0.0,~~~-0.0075)&1
000&4&0.0&0.98\\
\hline
&&& &&&&0.000079&0.98\\
\hline
&& &&&&&0.00016&0.97\\
\hline
&& &&&&&0.00032&0.87\\
\hline
&& &&&&&0.00064&0.68\\
\hline
\end{tabular}
\caption{CNOT sweeps with a shortened \ad time, using $\beta=1.14$.
``Final overlap'' is
shown for increasing noise parameter $\Delta$.
Due to the  increased
$\omega_{tunnel}$ permitting faster sweeps, a much stronger \den,
compared to $\beta=1.19$,
can be tolerated before the sweeps become unsuccessful. Other
parameters are as in Fig.\,14.  The noise frequency was
$\omega_c=0.042$. The entry with $\Delta=0.00032$ would
correspond to a \de time of 500 time units, or 3.1 ns with our
``typical'' parameters. }
\end{table}
\end{center}

\section{Conclusions}
 
We have presented a  description  of  one or two  interacting
rf \sqn s  in a form suitable for numerical simulation 
and a  \sy of programs for carrying out these simulations.
In addition to static properties, the behavior under time
dependent  conditions,  including flux  sweeps for \ad gates,
 and  the effects of  noise or \den, were studied.

 Among the points investigated was the validity of the ``spin 1/2
picture'' where the behavior governed by the full \h is
approximated by the components of an effective  ``spin 1/2''\sy 
using the  lowest levels of the double-potential well. The
identification  between the parameters
of the spin \h and the full \h was established and the regime of
validity of the approximation were investigated 
numerically. It is found, using various criteria, including
``Hilbert space
completeness'', that the ``spin
1/2'' picture is approximately valid up to  variations of the
external bias which  moves the basic pair of levels a substantial
fraction of the principal level splitting.

For the two \sq \syn, we have found  it is capable of performing  
the logical operation CNOT, and
 have been able to determine regions of the \pl plane with
definite configurations of the \wvs suitable for  CNOT sweeps.
The conditions for \ad behavior under such sweeps 
were  examined and \ad times  found for typical \sq
parameters.
The splitting $\omega_{tunnel}$ of the basic pair  enters into the
\ad condition and is very sensitive to the barrier parameter
$\beta$. It is found that the most interesting regions for this
parameter are values slightly greater than one, and detailed
studies
were presented for $\beta=1.19$.

Noise or \de were simulated as a random flux noise and a formula 
was derived relating the parameters of this random noise to the \de
parameter D. Numerical simulations verify the validity of the
formula. This random noise is then applied to the \ad logic gates 
and possible regimes of operation for the gates are found.
One of the interesting conclusions  is that the \ad sweep time 
can be substantially longer than the \de time 1/D before \de
effects become very large. This is traced to the fact that \den,
and also non-\adyn, are mostly effective at ``level crossing'',
which
is only a small part of the sweep. Indeed for the two-\sq CNOT gate
it is found that  noise applied to the control bit \sqn, which
does not undergo a ``level crossing'', has almost no effect. 

One of our general conclusions therefore,   for any devices of this
general type, is that ``level crossings'' should be held to a
minimum, and   when they occur
 a large tunneling energy or
splitting at ``crossing'' is beneficial both for reducing both \de
and non-\ady effects.

It thus appears that
an interesting region for the engineering of  \ad gates 
involves rather  small values of $\beta$, where the splitting is
large.
 However the \wvs are then nearly
delocalized. We have  examined the case
$\beta=1.14$, where
the reduced tunnel barrier and thus large $\omega_{tunnel}$
permit  fast operation of the gates. The resulting short time
for the \de to act leads to operations which are successful with
presently discussed values of the \de time, some nanoseconds.
However,  a careful choice of parameters is necessary.  

Generally speaking,
we may  say that our studies verify that it is in
principal possible to have a regime  where one can create and
manipulate quantum linear combinations of  quasi-classical objects,
as pictured in Fig.\,9. 

While the  parameters and examples we have used
are specific to the rf \sqn, it will be recognized that our methods
and results will apply, after some adjustments, to almost all
interacting double-potential well systems.


\begin{thebibliography}{00}
\bibitem{correct} We express frequencies  as angular frequency,
radians/second, since this what naturally arises  with $\hbar =1$
units.  To get Herz, divide by $2\pi$. We take this opportunity to
correct a
misunderstanding in refs~\cite{pla,ref1,deco} where frequencies
were
expressed as Hz: the units should have been
radians/second.  
  Other quantities, not cited as  Hz, need not be changed.
\end{thebibliography}
\end{document}